\documentclass[a4paper,twocolumn,superscriptaddress]{revtex4}

\usepackage{bm,amsmath}
\usepackage{graphicx}
\usepackage[normalem]{ulem}

\usepackage{xcolor}
\usepackage{hyperref}
\hypersetup{
     colorlinks,
     linkcolor={red!90!black},
     citecolor={black!10!blue},
     urlcolor={blue!80!black}
     }

\newcommand{\g}{\gamma}

\newcommand{\f}{\frac}
\newcommand{\p}{\partial}

\begin{document}

\title{Affinity and valence impact the extent and symmetry of phase separation of multivalent proteins}

\author{Saroj Kumar Nandi}
\affiliation{Department of Chemical and Biological Physics, Weizmann Institute of Science, Rehovot, Israel}
\affiliation{TIFR Centre for Interdisciplinary Sciences, Tata Institute of Fundamental Research, Hyderabad - 500046, India}

\author{Daniel {\"{O}}sterle}
\affiliation{Department of Chemical and Structural Biology, Weizmann Institute of Science, Rehovot, Israel}

\author{Meta Heidenreich}
\affiliation{Department of Chemical and Structural Biology, Weizmann Institute of Science, Rehovot, Israel}

\author{Emmanuel D. Levy}
\affiliation{Department of Chemical and Structural Biology, Weizmann Institute of Science, Rehovot, Israel}

\author{Samuel A. Safran}
\affiliation{Department of Chemical and Biological Physics, Weizmann Institute of Science, Rehovot, Israel}

\begin{abstract}
Biomolecular self-assembly spatially segregates proteins with a limited number of binding sites (valence) into condensates that coexist with a dilute phase. We develop a many-body lattice model for a three-component system of proteins with fixed valence in a solvent. We compare the predictions of the model to experimental phase diagrams that we measure \textit{in vivo}, which allows us to vary specifically a binding site's affinity and valency. We find that the extent of phase separation varies exponentially with affinity and increases with valency. Valency alone determines the symmetry of the phase diagram.

\end{abstract}
\maketitle

{\em \textbf{Introduction}:} Protein self-assembly plays a central role both in health and disease \cite{zbinden2020}. Aberrant protein assembly is responsible for neurodegenerative diseases such as Alzheimer’s or Huntington’s  \cite{peskett2018,ghosh2019}, type-II diabetes \cite{pytowski2020}, and others. Conversely, protein self-assembly into biomolecular condensates can spatially localize biochemical processes in membrane-less, mesoscale compartments \cite{banani2017,lyon2021}. These biomolecular condensates (BMCs) can comprise ribonucleic acids (RNAs), nucleic acids, and various proteins \cite{banani2017,lyon2021,shin2017,berry2018}. Examples of such BMCs include germline-granules \cite{smith2016,saha2016}, stress (responsive) granules \cite{protter2016,kucherenko2018,jayabalan2016,seguin2014,molliex2015,riback2017}, as well as chromatin-bound condensates \cite{strom2017}. BMCs also display crucial regulatory roles in various biological processes \cite{banani2017,mitrea2016,zbinden2020} such as cell differentiation \cite{liu2020,quiroz2020}, centrosome assembly \cite{zwicker2014,zwicker2015}, reaction kinetics \cite{li2012}, noise buffering \cite{klosin2020,stoger2016,dan2021}, or metabolic control \cite{prouteau2018}. Thus, general theoretical frameworks that enable conceptualization and prediction of the behavior of BMC systems are critical for comparison with experiments to foster their control. These include understanding concentrations, molecular interactions and temperature \cite{brangwynne2009,mitrea2016,feric2016,banani2017} under which biomolecules condensate instead of remaining dispersed, or assesing  whether the formation of a BMC is a quasi-equilibrium process \cite{hyman2014} or whether it is driven by biochemical reactions \cite{zwicker2014,zwicker2015}.

The analysis of experiments on \textit{in-vivo} cellular condensates is often inconclusive since one does not know all the molecular species involved nor the interactions among them. Such uncertainty limits the ability to  quantitatively predict the properties of phase separation of cellular condensates. For example, BMCs often comprise intrinsically disordered proteins \cite{banani2017} for which the structural valency (number of binding sites available per molecule), the effective valency (number of sites that are sterically, simultaneously accessible for inter-molecular contacts) and the interaction energy (affinity) between binding sites are typically unknown. In that respect, synthetic systems help bridge this gap, as their parameters are known by design. Indeed, the use of synthetic systems \textit{in-vitro} enabled understanding the impact of protein concentration, as well as interaction affinity on phase separation \cite{li2012,alberti2019}. We have developed a unique,  synthetic system based on interacting dimers and tetramers which enforces inter-molecular contacts, so that the structural valency is equal to the effective valency \cite{metapaper}. The modularity of this system also allows one to vary the valency parameter, for example replacing tetramers with hexamers, and changing the bond affinity, as we demonstrate in this work.

Additionally, BMCs fundamentally differ from systems studied in physics and physical chemistry, where one usually considers isotropically interacting, small molecule mixtures. In contrast, biomolecules are large, and in many cases, interactions are constrained by their geometry and the number and position of binding sites at their surface. Simple models with non-specific interactions predict phase separation with a critical volume fraction of the order of $1/2$ \cite{sambook,dillbook}. In contrast, for polymeric systems, Flory-Huggins theory \cite{degennesbook} predicts a critical concentration that scales as $N^{-1/2}$, where $N$ is the polymerization index. For large $N$, the system phase separates at very low polymer volume fractions with a fractal-like polymeric ensemble. In addition, for branched polymers, the classical theory of gelation by Flory and Stockmayer \cite{florybook,stockmayer1943,stockmayer1944} gives the percolation threshold of the network, a geometric property. However, our interest here is on the thermodynamic phase separation that leads to condensate formation. Reference \cite{zilman2003} presents the effects of varying multivalency on both phase separation and percolation. The latter, in particular, highlights that valence is a critical parameter. Our theory complements the existing approaches listed above: we focus on the primary physical mechanisms relating phase separation to finite multivalency of rigid proteins. Progress along this direction has come from molecular-dynamics simulations of patchy particles \cite{bianchi2006,smallenburg2013}, the extension of Wertheim theory \cite{wertheimI,wertheimII,wertheimIII,wertheimIV}, and lattice-based Monte-Carlo simulations of model biological proteins \cite{harmon2017,harmon2018,choi2019}. In this work, we develop a relatively simple lattice model that automatically includes the excluded volume effect and the multi-body nature of finite valence. This makes the theory amenable to analytical treatment and facilitates comparison with experiments on the extent of phase separation as a function of affinity and valence.

\begin{figure}
 \includegraphics[width=8.6cm]{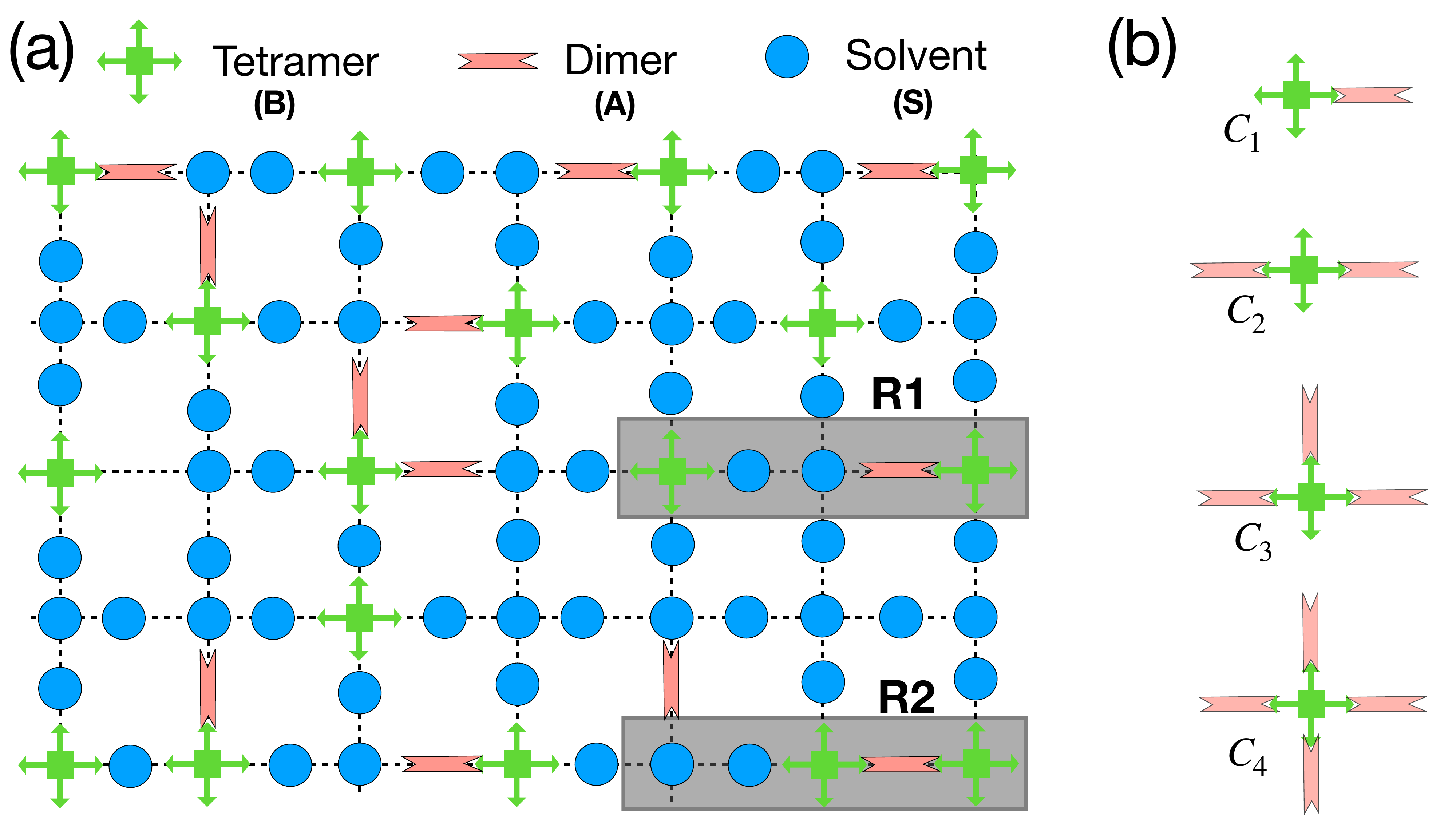}
 \caption{Schematic description of the lattice model for the solution of tetramers, dimers and solvent. (a) Lattice model for the system on a square lattice. Regions R1 and R2 show the same particles in two different configurations:  the configuration in R2 has higher enthalpy owing to satisfied interactions between $A$ and $B$ and a lower entropy as other configurations of $A$ and $B$ are restricted by the interaction. The configuration in R1 has lower enthalpy and higher entropy than R2. (b) Four possible complexes when the $B$ molecules are tetramers.}
 \label{latticemodel_schematic}
\end{figure}

\begin{figure}
	\includegraphics[width=8.6cm]{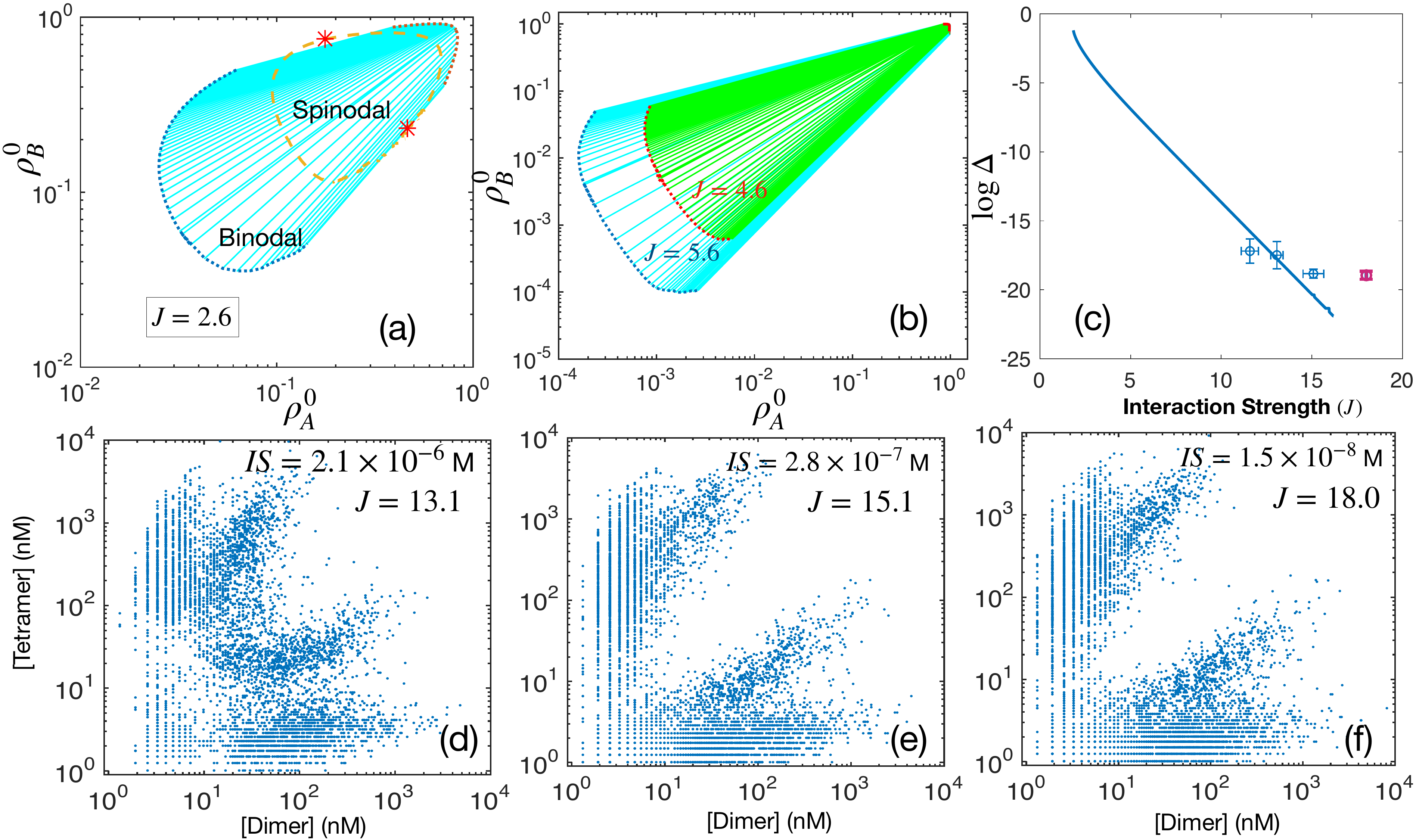}
	\caption{(a) Phase diagram for a solution of tetramers, dimers and solvent. The dotted line denotes the binodal and the end points of the tie-lines (in light blue) are the concentrations of the concentrated (dashed) and dilute (dotted) phases. The yellow dashed line is the spinodal that gives the limit of metastability. The two red stars denote the two critical points where the tie line length vanishes and the two coexisting phases become identical. (b) Binodal for solutions of tetramers, dimers and solvent at two different values of the interaction strength, $J$, show stronger phase separation (larger area of the two-phase region and larger differences in concentrations of the two coexisting phases) at larger $J$. (c) The minimum distance $\Delta$ of the phase boundary from origin as a function of interaction strength $J$. The line is the theory and the symbols (with error bars) are experimental data. We note that the datapoint for the highest interaction strength (purple) overestimates $\Delta$ because the corresponding concentrations of $A$ and $B$ in the dilute phase reach a value below the detection limit of the microscope (d-f) Experimental phase diagrams for the dimer-tetramer system with different interaction strength (affinities) ($J=-\log(IS)$ in units of $k_BT$).}
	\label{phasedia_valence4}
\end{figure}

\begin{figure}
	\includegraphics[width=8.6cm]{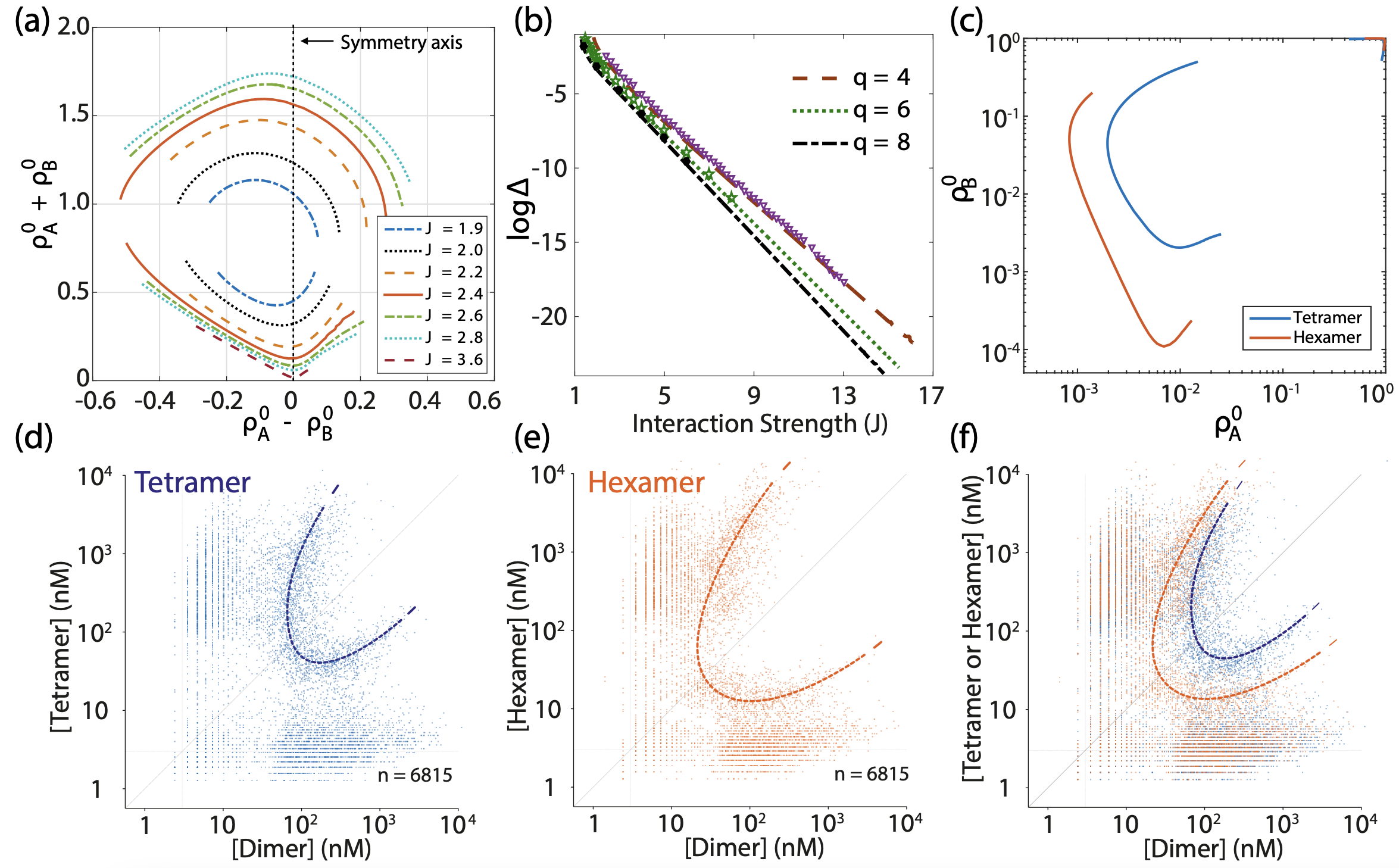}
	\caption{ (a) The symmetry axis of the phase diagrams for phase separation of tetramers and dimers lies along the zero of the abscissa for asymptotically large $J$. This shows that phase separation is strongest at the stoichiometric ratio (of the effective concentrations).  (b) Comparison of the value of the minimal concentration, $\Delta$, obtained from the full model (points) and simplified model (lines) along the symmetry axis. (c) Theoretical phase diagram for both tetramers/dimers and hexamers/dimers both with $J=4.0$ shows stronger phase separation for the hexamers. (d-f) Experimental phase diagrams for the same interaction strength and different valence (d) Phase diagram for a dimer-tetramer system (e) Phase diagram for a dimer-hexamer system (f) Experimental data from panels (d) and (e) are overlaid. $[\ldots]$ denotes concentration of interaction sites and dotted lines are approximate phase boundaries.}
	\label{mindist_theory_expt}
\end{figure}

Our focus here is the phase separation into concentrated and dilute regions in a three-component system of a solvent and two proteins: one of which is divalent (dimer) and another with a valence larger than two (multimer). The multimers interact among themselves via the dimer, which links two multimers as schematically shown in Fig. \ref{latticemodel_schematic}. The geometric design of the proteins prevents intra-molecular binding, i.e., where two sites of a dimer bind on the same multimer. 
Experimentally, we genetically encode such a pair of proteins and monitor their expression and phase separation in yeast cells, as described in Ref. \cite{metapaper}. Briefly, the dimer and multimer each consist of three structured domains fused by flexible linkers: the first domain is a fluorescent reporter, the second confers multivalence by homo-oligomerization, and the third mediates the affinity or interaction strength ($IS$) between the dimer and the multimer. Uniquely, both $IS$ and the multimer valency can be modulated by the experimental, molecule design. The dimer and tetramer are coexpressed in the cytoplasm of yeast cells and undergo phase separation at high enough concentrations. Protein concentration in the dilute phase is quantified by fluorescence microscopy and is compared with our theory. Theoretically, we predict the phase diagram topology and symmetry (the axis of maximum phase separation) for the association of such multimers linked (or not) by dimers. We find that the phase boundaries enclosing the coexisting regions form closed loops and crucially depend on the valence and relative affinity, $IS$, between the dimers and multimers (Fig. \ref{phasedia_valence4}). We, therefore, focus on the phase diagrams as a function of the dimer and multimer concentrations at various interaction strengths (Fig. \ref{phasedia_valence4}). For these coexistence curves, the valence determines the symmetry that depends on the multimer-dimer ratio (Fig. \ref{mindist_theory_expt}b); this is different from closed-loop phase diagrams in the temperature-solute concentration plane for hydrogen-bonding systems \cite{wheeler1980,walker1980}. In addition, our theory predicts that the minima (minimum distance from the origin), $\Delta$, of the phase diagrams, vary exponentially with interaction strength, which is in agreement with experimental data (Fig. \ref{phasedia_valence4}c-f). 
We then elucidate the role of multivalency in phase separation and how valency affects the rate of decrease of the distance to the origin $\Delta$ with $IS$ (Fig. \ref{mindist_theory_expt}c). Finally, we show within the theory and the experiment that phase separation becomes more effective (i.e., phase boundaries cover a larger region of the concentration space) at higher valence for a fixed interaction strength (Fig. \ref{mindist_theory_expt}d-f).

{\it \textbf{Theoretical model}:}
We formulate a statistical mechanical description of the system of multimers and dimers, which we solve within a mean-field approximation. Since the focus of the experiments is on the topology and symmetry of the phase diagrams as a global function of the two concentrations at fixed temperature, the corrections to mean-field theory are important only near the critical points and therefore are not of interest here. Instead, we focus on the experimentally important parameters of compositions, valence, and interaction strengths. We designate the dimers by $A$ and the multimers by $B$. The proteins in the experiments are designed so that $AB$ interact as lock and key \cite{li1998,metapaper}. Additionally, two interaction sites of $A$ cannot bend to interact with two sites of the {\em same} $B$ molecule due to the rigidity of A. Therefore, phase separation proceeds through intermolecular associations between $A$ and $B$.

For concreteness, we first  consider a particular example: a dimer and a tetramer being the $A$ and $B$ particles, respectively, and the rest of the system is considered as a uniform (mostly aqueous, in the case of a cell), solvent $S$.
From the experiment, we find the phase separation to be strongest (i.e., the concentration difference between the two coexisting phases is largest) at the stoichiometric ratio of interaction sites (volume fraction of molecules multiplied by their valence) of $A$ and $B$.
To elucidate this within our mean-field theory, we consider a lattice model where the $A$ molecules occupy only the bonds and the $B$ molecules occupy only the sites of the lattice. Solvent molecules, $S$, can occupy either the bonds or sites as schematically shown in Fig. \ref{latticemodel_schematic}(a). $N_s$ and $N_b$ denote the total number of sites and bonds, respectively.
Since the $B$ molecules have four interaction sites ($q=4$) each, we consider a square lattice, where $N_b = 2N_s$; to treat other valences, $q$, we use different lattices (see SM, Sec. V).
The system contains a total of $N_0^A$ $A$ molecules and $N_B^0$ $B$ molecules. Since the lattice is fully occupied, conservation dictates that there must be $(2N_s-N_A^0)$ $S$ molecules on the bonds and $(N_s-N_B^0)$ on the sites.

Modeling the experimental phase diagrams requires the inclusion of many-body interactions to account for the finite valency, even in mean-field theory \cite{radhakrishnan1999,radhakrishnan2005}. To do so, we proceed in two separate stages: first, the $A$ and $B$ molecules associate with each other forming complexes, and second, the complexes interact among themselves as well as with the free $B$ molecules (i.e., those not associated with any $A$) leading to phase separation. 
To simplify the problem and obtain physical insight, we use a mean-field approximation where the complexes interact with the average concentration of $B$ molecules.
For the particular case of tetramers and dimers, there can be four different complexes: $C_i$ with $i = 1, 2, 3, 4$, where $C_i$ denotes a configuration with $i$ $A$ molecules associated with one $B$ molecule as schematically shown in Fig. \ref{latticemodel_schematic}(b). 

To illustrate the physical origin of the phase separation, consider the two shaded regions $R1$ and $R2$, in Fig. 1(a): they both contain the same number of particles, one $A$, two $B$, and two $S$. When the attractive interaction dominates, the configuration in $R2$ has lower free energy compared to that in $R1$; in contrast, when entropy dominates, the arrangement in $R1$ has lower free energy than that in $R2$. In equilibrium, the system configuration is that which minimizes its free energy:  When the enthalpy term dominates, it favors $R2$, and the system phase separates; on the other hand, when the entropy term dominates, it favors $R1$, and the system remains in a homogeneous, single phase.

After the complexes have formed, the dimensionless concentration of free $A$ molecules (the fraction of bonds occupied by uncomplexed $A$ molecules) is $\rho_A = N_A /N_b$, where $N_A$ is the number of free $A$ molecules. Similarly, $\rho_B = N_B/N_s$ is the dimensionless concentration of free $B$ molecules. This particular normalization uses the effective concentrations of the interaction sites (i.e., actual concentrations multiplied by valence), which is the quantity that we also use for analysis of the  experiments.
The effective  concentrations of the total (overall) $A$ and $B$  interaction sites are $\rho_A^0$ and $\rho_B^0$, respectively, and the concentrations of the $i$th complex are $\gamma_i$. Then, the total free energy (see supplementary material (SM), Sec. II for detail), $f$, per site, in units of $k_BT$, where $k_B$ is the Boltzmann constant and $T$, the temperature, is
\begin{align}\label{freeenergy}
f =&2\rho_A\ln\rho_A+\rho_B\ln\rho_B+2(1-\rho_A^0)\ln(1-\rho_A^0) \nonumber\\
&+(1-\rho_B^0)\ln(1-\rho_B^0)+\gamma_1\ln(4\g_1)+\g_2\ln(6\g_2) \nonumber\\
&+\g_3\ln(4\g_3)+\g_4\ln\g_4 \nonumber-J(\g_1+2\g_2+3\g_3+4\g_4)\\
&-(J-J_{BB})\rho_B^0(\g_1+2\g_2+3\g_3+4\g_4)
\end{align}
where the product $iJ$ is the gain in binding energy (in units of $k_BT$) due to the formation of $C_i$. $J_{BB}$ is a parameter governing the change in interaction when both sides of the dimer, compared to only one of its sides, is attached to  the corresponding site on the $B$ molecules. Here we consider $J_{BB}=0$ and comment on non-zero $J_{BB}$ in SM, Sec. IX. Note that we have treated the solvent on the sites and the bonds as two different states, since the volumes occupied by $A$ and $B$ molecules can be different.\\
\\
{\it \textbf{Modeling the effect of interaction strength on phase separation}}

Conservation of the $A$ and $B$ molecules respectively implies that $\rho_A=\rho_A^0-(\g_1-2\g_2-3\g_3-4\g_4)/2$ and $\rho_B=\rho_B^0-\g_1-\g_2-\g_3-\g_4$ where the $\gamma_i$ are the concentrations of the complexes. For a given  interaction strength $J$, we first minimize $f$ with respect to $\g_i$'s; this leads to four equations which we solve simultaneously to obtain the $\g_i$'s in terms of $\rho_A^0$ and $\rho_B^0$, which then allows us to calculate the phase diagrams (see SM, Sec. II). The phase diagrams are functions of the interaction strength $J$, $\rho_A^0$ and $\rho_B^0$, and therefore are three-dimensional. 
We  plot the phase diagrams in the two-dimensional plane of the given effective concentrations, $\rho_A^0$ and $\rho_B^0$, at a fixed value of $J$. There is no phase separation at small values of $J$; as $J$ increases and crosses a certain value (depending on valence), phase separation occurs. We find a closed-loop phase diagram with two critical points as shown for $J=2.6$ in Fig. \ref{phasedia_valence4}(a).
The outer boundary, marked by the dotted line, is the binodal, and the dashed line is the spinodal. Phase separation occurs inside the binodal region where dense (red) and dilute (blue) regions coexist in equilibrium, whereas outside this boundary, the system remains homogeneous. The two stars mark the critical points where the concentrations of dense and dilute regions become identical. The lines connecting the dense and the dilute regions are the tie-lines, whose end-points are the effective concentrations of two coexisting regions. If the interaction strengths and overall concentrations are such that the  effective concentrations are within  the spinodal region, the uniform, mixed state is unstable, and phase separation occurs via spinodal decomposition along the tie lines. On the other hand, if the system lies between the spinodal and the binodal regions, phase separation proceeds through nucleation and growth \cite{bray1994,sambook,zilman2003}.

Figure \ref{phasedia_valence4}(b) shows the binodal phase diagrams for the tetramer-dimer system at two different values of $J$; increasing $J$ leads to more effective phase separation (as defined above). 
For a quantitative comparison of theory and experiment, we define $\Delta$, the minimum distance of the dilute region of the binodal, as a function of different $J$ and the corresponding experimentally varied affinity. In a two-component system, $\Delta$  decreases exponentially with $J$ when $J\gg 1$(see SM, Sec. VI); we expect and observe a similar behavior for the three-component system. We plot $\Delta$ derived from the numerical solution of the theory (line) and from the experiment (symbols, see SM, Sec. I S3) in Fig. \ref{phasedia_valence4}(c). 
The experimental uncertainty in both the affinity (corresponding to the interaction strength, $J$) and the measured effective concentrations are shown in the figure. Note that the concentrations at the largest affinities in experiments are at the limits of our experimental resolution. The agreement in Fig. \ref{phasedia_valence4}(c) is close given the experimental uncertainties. We also show the experimental phase diagrams at three different affinities (interaction strengths $IS$, measured in units of $M$, where $J=-\log(IS)$ in units of $k_BT$). The points in Figs. \ref{phasedia_valence4}(d-f) are the effective concentrations of interaction sites of dimers and tetramers for cells that do not exhibit a visible condensate in them; thus, the data from the many cells with different protein concentrations depicts the part of the binodal that delineates the dilute phase \cite{metapaper}.

Although the theoretical phase diagram is a closed-loop, a quantitative measurement of the concentration in the dense phase is challenging due to the limited axial resolution of the microscope, inner filter effects, and foster energy transfer. Therefore, we compared only the dilute region of the measured phase diagrams with the theory. In the experimental system, $A$ and $B$ interact with an affinity on the order of 100 $nM$ or $\sim 15 k_BT$ \cite{metapaper}. The numerical solution of the theory at such  affinities is impractically slow (see SM, Sec. VII) and we did not calculate the entire phase diagram. However, we discuss in the SM and show in Fig. \ref{mindist_theory_expt} that the extreme regions of the phase diagram (those that connect largest tie lines) can be calculated even for large values of $J$. We saw in Fig. \ref{phasedia_valence4} that the shapes of the phase diagrams for the dilute region, and the trend with increasing $J$, are similar to the situation at smaller values of $J$. When the affinity is high, the system will use more $A$ molecules to associate the $B$ molecules, as schematically shown in Fig. \ref{latticemodel_schematic}(a). 
As shown in the SM (Fig. S3), when the affinity of $A$ to $B$ is weak, the likelihood of complex formation is low, and most of $A$ molecules are free; whereas, for high affinity, the probability is high, and most of the $A$ molecules are incorporated into  complexes. These complexes then interact with the free $B$ particles or other complexes to yield the dense phase.

{\it \textbf{Effect of valency}:} We expect maximal phase separation when the effective concentrations of interaction sites for the two species are equal. To test this hypothesis, we plot the phase diagrams for a mixture of dimers and tetramers at several values of $J$ as functions of $\rho_A^0 -\rho_B^0$ vs $\rho_A^0+\rho_B^0$ in Fig. \ref{mindist_theory_expt}(a). 
As $J$ increases, we expect the symmetry-line, i.e., the tie-line of maximal phase separation, to lie on the zero of the abscissa, which we indeed observe (Fig. \ref{mindist_theory_expt}(a)).
If stoichiometry determines the maximal extent of phase separation, $\Delta$ should lie along this symmetry axis. To test this hypothesis, we approximate $\rho_A^0=\rho_B^0$ and solve the phase diagrams along this symmetry axis (see SM, Sec. VIII) $\Delta$ obtained using this simple  approximation agrees well with the numerical results of the complete theory (Fig. \ref{mindist_theory_expt}b).

We now discuss the role of multi-valency. For concreteness, we consider two different systems:  tetramers or hexamers, both with dimers and solvent (see SM, Sec. V).
All other parameters being equal, we find that the system with hexameric $B$ molecules shows a larger region of phase separation, compared with tetramers, as shown in Fig. \ref{mindist_theory_expt}(c) for $J=4.0$. To compare the theory with experiments, we measured the phase diagrams for both systems: tetramers or hexamers with dimers; the phase diagrams are shown in Figs. \ref{mindist_theory_expt}(d-f). Consistent with  theory, phase separation is more effective for the hexamers than the tetramers at a given interaction strength. Note that the trivial factor that accounts for a larger number of interaction sites per multimer is accounted for in these plots as we consider concentrations of interaction sites, not concentrations of $B$ molecules. The system with larger valence shows stronger phase separation due to the availability of more complexes, which increases the interaction energy per particle.
Smaller values of $\Delta$ with increasing multivalency $q$ at a fixed $J$ (Fig. \ref{mindist_theory_expt}c) are also consistent with the experimental results.

In summary, we have presented a  simple theory that predicts phase separation in a three-component system of multivalent proteins: where one of the components is dimeric, and the other has a higher valence.
While we have shown the results for tetramers, hexamers, and octamers, the theoretical approach can be extended to other systems.
The theory is motivated by and compared with experiments on cytoplasmic phase separation within yeast cells where the phase separating proteins are synthetic and foreign to those cells \cite{metapaper}. Since these proteins are not expected to interact with the intrinsic proteins of the cells, the experimental system allows quantitative control over the interaction strengths and valency of the system compared with cellular protein condensates which can have any number of  additional, unknown components. In most cases of intra-cellular phase separation, the details of protein-protein interactions are unknown; our experimental system, along with the analytical theory, should be viewed as a step towards a quantitative understanding of the phase separation process {\it in-vivo}.

SKN would like to thank Ohad Cohen, Dan Deviri, Timon Idema and A. Radhakrishnan for discussions and the Koshland Foundation for funding through a fellowship. SAS is grateful for the support of the Israel Science Foundation, the Krenter and Perlman Family Foundations, the Volkswagen foundation,  and a Katz-Krenter Center grant. EDL acknowledges support by the Israel Science Foundation (1452/18), by the European Research Council (ERC) under the European Union’s Horizon 2020 research and innovation programme (grant agreement No. 819318), by a research grant from A.-M. Boucher, by research grants from the Estelle Funk Foundation, the Estate of Fannie Sherr, the Estate of Albert Delighter, the Merle S. Cahn Foundation, Mrs. Mildred S. Gosden, the Estate of Elizabeth Wachsman, the Arnold Bortman Family Foundation.

%


\onecolumngrid
 \subsection*{\underline{Supplementary Material}\\ Affinity and valence impact the extent and symmetry of phase separation of multivalent proteins}
 
  In this supplementary material, we provide a summary of the experimental system and the details of the calculation for the mean-field lattice free energy for a solution of tetramers, dimers and solvent. We then discuss how to obtain the phase diagrams, the roles of the complexes and the specific details of the comparison of two mixtures, one consisting of hexamers, dimers and solvent and the other of tetramers, dimers and solvent, to elucidate the role of valency in the phase separation process. Finally, we discuss the behavior of the minima of phase diagram in a simple two-component system, a summary of the numerical solution and the simplified model along the symmetry axis.
  
  \vspace{1.5cm}
  \twocolumngrid

  \setcounter{table}{0}
  \setcounter{figure}{0}
  \setcounter{section}{0}
  \setcounter{equation}{0}
  \renewcommand{\thetable}{S\arabic{table}}
  \renewcommand\thefigure{S\arabic{figure}}
  \renewcommand{\theequation}{S\arabic{equation}}
  \renewcommand{\thesection}{S\arabic{section}}

  \maketitle
  \begin{figure*}
  	\includegraphics[width=16cm]{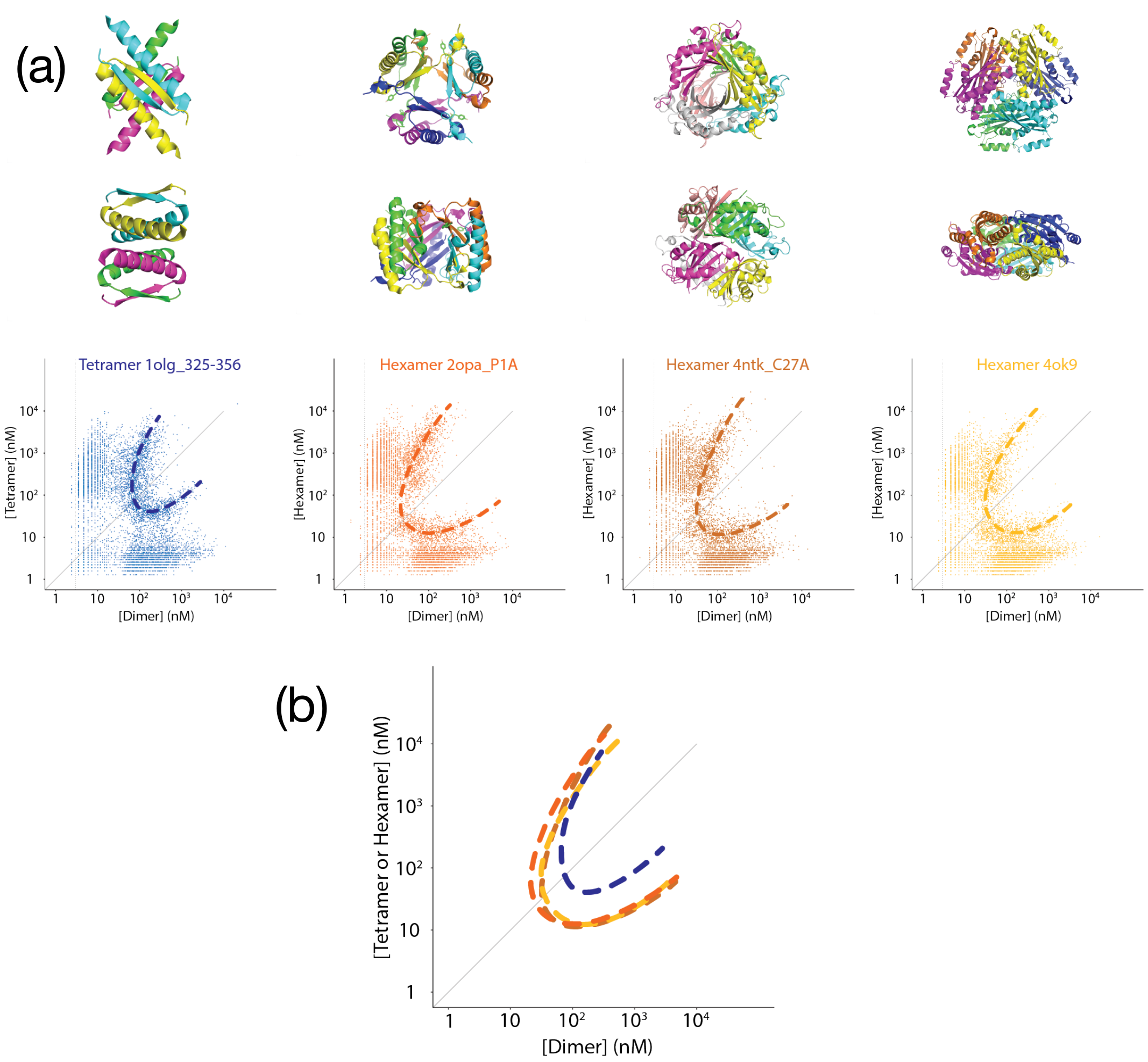}
  	\caption{(a) Structures of the different oligomeric scaffolds and their associated experimental phase diagram shown underneath. The overlay of all phase boundaries highlights consistency within scaffolds with six binding sites (hexamers) and a clear shift as compared to the reference scaffold with four binding sites (tetramer). The 4-letter PDB code corresponding to each scaffold is given above its respective phase diagram. 
  		(b) $\Delta$ is obtained  manually from an estimate of the most probable phase diagram in each case. The other two lines provide the extent of maximum error in the estimation of $\Delta$.}
  	\label{exp_details}
  \end{figure*}
  
  \begin{figure*}
  	\includegraphics[width=17.4cm]{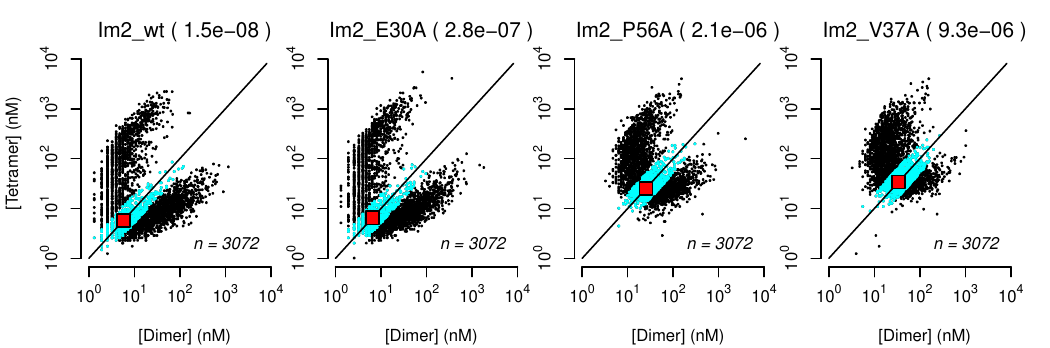}
  	\caption{ The lowest concentration of the phase boundary was determined as the peak density (red square) of selected data points (cyan) along the diagonal. Affinities of different Im2 mutants decrease from left to right and accordingly the lowest point of the phase boundary shifts to higher concentrations. 
  		$\Delta$ represents the distance of the red square from the origin. 
  		The limit of resolution to measure concentration in our fluorescence microscope is about $6nM$, which means that $\Delta$ might be smaller than the value observed for the affinity $IS=1.5\times 10^{-8}M$.}
  	\label{exp_details_delta}
  \end{figure*}

  \section{Details of the experimental system}
  The synthetic, two-component protein system used in the experimental work has been  previously described in Ref. \cite{metapaperSM}. Briefly, a first component consists of a large homodimer that we designate $A$. The homodimer is fused to a red fluorescent reporter at its C-terminus and a binding domain Im2 (Immunity protein 2) at its N-terminus. We designate $B$ the second component, which consists of a yellow fluorescent protein at the N-terminus, a binding domain E9 (E. coli Colicin 9, which binds to Im2 specifically) and a homo-oligomerization domain at the C-terminus that can be a tetramer ($q=4$) or a hexamer ($q=6$). The large size of the dimer and small size of the tetramer/hexamer prevent intramolecular contacts (i.e., two sites of the dimer binding to two sites on the same tetramer/hexamer). Such intramolecular contacts would inhibit the phase separation of the system. Here, we used the wild type version of the Im2 domain as well as mutants E30A, P56A, and V37A, which affinities to E9 have been previously measured  (PMID: 32661377). Thus, employing those mutants allowed us to control the interaction strength ($IS$) between the dimer and tetramer/hexamer. $IS$ in the experiment is measured in units of $nM$ and related to $J$ (in units of $k_BT$, the Boltzmann constant times the temperature) by $J=-\ln(IS/c^0)$, where $c^0=1M$ is the reference concentration.
  
  For valence-variation experiments the dimer-tetramer system served as a reference. Increasing the valency of the interacting components is expected to increase their effective affinity. Therefore, we based these experiments on a low affinity mutant of Im2 (V37A) to enable detecting further increases in affinity associated with changes in valence. In order to change the valence, the tetramerization domain was replaced with selected homo-oligomerization domains from thermophilic organisms, as these tend to exhibit high protein stability. The main selection criteria for homo-oligomerization domains was a small size and a validated homo-oligomeric state as verified through PiQSi and QSbio (PMID: 17997962 \& 29155427). Candidates with four, six, eight, ten, and twelve binding sites were chosen. From these initial candidates, those with six binding sites (hexamers) showed consistent phase boundaries despite all hexameric scaffolds exhibiting different tertiary structures (Supplementary Fig. 1), implying that the origin of the change in the phase boundaries came specifically from the change of valence. In contrast, the phase diagrams of other valencies did not show consistent phase boundaries across multiple structures (10, 12, 24-mers) nor an enhanced phase separation relative to the tetramer (8-mers).  Such inconsistent phase boundaries could be caused by several confounding factors such as the presence of intramolecular interactions, steric hindrance preventing a regular pavement in 3D space, improper folding, or improper assembly due to expression in a heterologous organism. For these reasons, it was difficult to interpret our observations for valencies eight-and-up, and we focused on hexamers in this work.
  Detailed information about the scaffolds’ layout and sequences can be found in Supplementary Table 1-2.
  
  \subsection{Plasmids and Strains}
  The creation of plasmids for interaction strengths  is described in (PMID: 32661377). 
  To create plasmids with novel oligomerization domains, the plasmid carrying the tetrameric protein scaffold from our previous work (PMID: 32661377) was onboarded with Twist Bioscience (San Francisco, CA) as a custom vector. Plasmids in which only the original tetramerization domain was replaced with selected candidates were then ordered and co-transformed into a diploid S. cerevisiae strain created by mating BY4741 and BY4742 strains (PMID: 9483801). Individual components were also transformed in haploid (BY4741) cells (PMID: 9483801) to measure their expression. The tRNA adaptation index of newly selected scaffold proteins was optimized using custom scripts.
  
  \subsection{Microscopy and image processing}
  
  Sample preparation and imaging were performed as described previously. (PMID: 32661377). Image processing was done as previously described in (PMID: 32661377). Briefly, custom scripts (bioRxiv: 260695) in ImageJ/FIJI (PMID: 22743772) were used for identification and segmentation of individual cells. Among other properties the median fluorescence intensity was recorded for each cell. Fluorescence intensity values were converted into concentration values  based on a linear model of fluorescence intensity measured at different concentrations of purified fluorescent proteins, as described in (PMID: 32661377).
  
  \subsection{Analysis of the minimal distance, $\Delta$, of phase boundaries from the origin }
  The impact of valency on phase separation was quantified by determining the point of lowest concentrations on the phase boundaries, which we define by the distance, $\Delta$, of this point from the origin. In this analysis, we only considered cells containing condensates as they lie on the phase boundaries, which makes it possible to estimate its position on the diagram. For each cell condensate, the concentration of both scaffolds in the dilute phase was extracted and log-transformed, yielding x- (concentration of dimer binding sites) and y-coordinates (concentration of tetramer/hexamer binding sites). Thereafter, data points within a certain distance ($<1$) from the diagonal were selected. These points were then projected onto the diagonal and the peak of their density was determined. The data point closest to the location of the peak was then taken as the minimal concentration of the phase boundary. Each strain was imaged several times, and this analysis was repeated for each replicate. The final value was determined as the mean of all replicates.
  Importantly, there is a limit to this estimate, of  about $\sim 6nM$, due to the resolution of the microscope used to measure the auto-fluorescence of cells. This means we cannot resolve values of $\Delta$ that are lower than this limit. Thus, when $J$ is very large, such that $\Delta$ is very low, our experimental resolution fails to correctly measure the concentrations. The phase diagram for $J\sim 18$ or $IS=1.5\times10^{-8}M$, shown in Fig. 2(f) in the main text, yields boundaries at or close to this experimental resolution limit.
  
  Note that the experimental phase diagrams (in Figs. 2 and 3 in the main text and Fig. \ref{exp_details}) contain some data points representing single-phase in the two-phase region. In an ideal scenario, these data points should not exist. But the experimental phase diagrams come from {\it in-vivo} measurements in living yeast cells, and there are many sources of noise. Given enough time, we believe that the proteins inside these cells may eventually undergo phase separation eliminating these data points.

  \begin{table*}
  	\caption{Plasmids used to vary the interaction affinity.
  		All plasmids are based on the p41x vector (PMID: 7737504). The interaction affinities corresponding to the mutants  were taken from the literature. (PMID: 9718299)
  	}
  	\label{tab:my-table}
  	\resizebox{\textwidth}{!}{%
  		\begin{tabular}{ccccc}
  			\textbf{Name} & \textbf{Valency} & \textbf{ORF}                                                                                                          & \textbf{Affinity to E9 (M)} & \textbf{Yeast Selection Marker}                            \\
  			pMH04         & 2                & GPD Promoter - NES - Im2 - 4LTB - FusionRed - CYC Terminator                                                          & 1.5e-8                      & Hygromycin                                                 \\
  			pMH05         & 2                & GPD Promoter - NES - Im2 (V37A) - 4LTB - FusionRed - CYC Terminator                                                   & 9.3e-6                      & Hygromycin                                                 \\
  			pMH06         & 2                & GPD Promoter - NES - Im2 (E30A) - 4LTB - FusionRed - CYC Terminator                                                   & 2.8e-7                      & Hygromycin                                                 \\
  			pMH08         & 2                & GPD Promoter - NES-Im2 (P56A) - 4LTB - FusionRed - CYC Terminator                                                     & 2.1e-6                      & Hygromycin                                                 \\
  			pMH01         & 4                & GPD Promoter - Venus - E9 - 1AIE - CYC Terminator                                                                     & -                           & G418                                                       \\
  			yDO02         & BY4743           & \begin{tabular}[c]{@{}c@{}}Condensate strain with\\ Im2 (V37A) and 4NTK (C27A) as oligomerization domain\end{tabular} & pMH05, pDOv602              & \begin{tabular}[c]{@{}c@{}}G418,\\ Hygromycin\end{tabular} \\
  			yDO03         & BY4743           & \begin{tabular}[c]{@{}c@{}}Condensate strain with\\ Im2 (V37A) and 4OK9 as oligomerization domain\end{tabular}        & pMH05, pDOv603              & \begin{tabular}[c]{@{}c@{}}G418,\\ Hygromycin\end{tabular}
  		\end{tabular}%
  	}
  \end{table*}

  \begin{table*}[t]
  	\caption{Plasmids encoding scaffolds with variable valencies. 
  		Plasmids pDOv601 and pDOv602 feature point mutations in the oligomerization domain (P1A and C27A, respectively), in order to inactivate enzymatic activity.
  	}
  	\label{tab:my-table}
  	\resizebox{\textwidth}{!}{%
  		\begin{tabular}{ccccc}
  			\textbf{Name} & \textbf{Valency} & \textbf{\begin{tabular}[c]{@{}c@{}}PDB Code\\ Oligo- merization\\ Domain\end{tabular}}                                & \textbf{ORF}                                                       & \textbf{Yeast Selection Marker}                            \\
  			pMH05         & 2                & 4LTB                                                                                                                  & GPD Promoter - NES - Im2 (V37A) - 4LTB - FusionRed - CYCterminator & Hygromycin                                                 \\
  			pMH01         & 4                & 1AIE                                                                                                                  & GPD Promoter - Venus - E9 - 1AIE - CYC Terminator                  & G418                                                       \\
  			pDOv601       & 6                & 2OPA (P1A)                                                                                                            & GPD Promoter - Venus - E9 - 2OPA (P1A) - CYC Terminator            & G418                                                       \\
  			pDOv602       & 6                & 4NTK (C27A)                                                                                                           & GPD Promoter - Venus - E9 - 4NTK (C27A) - CYC Terminator           & G418                                                       \\
  			pDOv603       & 6                & 4OK9                                                                                                                  & GPD Promoter - Venus - E9 - 4OK9 - CYC Terminator                  & G418                                                       \\
  			yDO02         & BY4743           & \begin{tabular}[c]{@{}c@{}}Condensate strain with\\ Im2 (V37A) and 4NTK (C27A) as oligomerization domain\end{tabular} & pMH05, pDOv602                                                     & \begin{tabular}[c]{@{}c@{}}G418,\\ Hygromycin\end{tabular} \\
  			yDO03         & BY4743           & \begin{tabular}[c]{@{}c@{}}Condensate strain with\\ Im2 (V37A) and 4OK9 as oligomerization domain\end{tabular}        & pMH05, pDOv603                                                     & \begin{tabular}[c]{@{}c@{}}G418,\\ Hygromycin\end{tabular}
  		\end{tabular}%
  	}
  \end{table*}

  \maketitle
  \begin{table*}[t]
  	\caption{Yeast strains used in this work}
  	\label{tab:my-table}
  	\resizebox{\textwidth}{!}{%
  		\begin{tabular}{ccccc}
  			\textbf{Name} & \textbf{Background} & \textbf{Description}                                                                                                  & \textbf{Plasmids} & \textbf{Selection}                                         \\
  			yMH05         & BY4743              & \begin{tabular}[c]{@{}c@{}}Condensate strain with\\ wild type Im2\end{tabular}                                        & pMH01, pMH04      & \begin{tabular}[c]{@{}c@{}}G418,\\ Hygromycin\end{tabular} \\
  			yMH06         & BY4743              & \begin{tabular}[c]{@{}c@{}}Condensate strain with\\ Im2 (V37A)\end{tabular}                                           & pMH01, pMH05      & \begin{tabular}[c]{@{}c@{}}G418,\\ Hygromycin\end{tabular} \\
  			yMH07         & BY4743              & \begin{tabular}[c]{@{}c@{}}Condensate strain with\\ Im2 (E30A)\end{tabular}                                           & pMH01, pMH06      & \begin{tabular}[c]{@{}c@{}}G418,\\ Hygromycin\end{tabular} \\
  			yEl09         & BY4743              & \begin{tabular}[c]{@{}c@{}}Condensate strain with\\ Im2 (P56A)\end{tabular}                                           & pMH01, pMH08      & \begin{tabular}[c]{@{}c@{}}G418,\\ Hygromycin\end{tabular} \\
  			yDO01         & BY4743              & \begin{tabular}[c]{@{}c@{}}Condensate strain with\\ Im2 (V37A) and 2OPA (P1A) as oligomerization domain\end{tabular}  & pMH05, pDOv601    & \begin{tabular}[c]{@{}c@{}}G418,\\ Hygromycin\end{tabular} \\
  			yDO02         & BY4743              & \begin{tabular}[c]{@{}c@{}}Condensate strain with\\ Im2 (V37A) and 4NTK (C27A) as oligomerization domain\end{tabular} & pMH05, pDOv602    & \begin{tabular}[c]{@{}c@{}}G418,\\ Hygromycin\end{tabular} \\
  			yDO03         & BY4743              & \begin{tabular}[c]{@{}c@{}}Condensate strain with\\ Im2 (V37A) and 4OK9 as oligomerization domain\end{tabular}        & pMH05, pDOv603    & \begin{tabular}[c]{@{}c@{}}G418,\\ Hygromycin\end{tabular}
  		\end{tabular}%
  	}
  \end{table*}
  
  \section{Free energy for the mixture of  tetramers,  dimers and solvent}
  \label{tetramerfreeenergy}
  We are interested in the phase separation  of a three-component system consisting of a linear molecule $A$ with two interaction sites, a multivalent molecule $B$ with valency (i.e., interaction sites) $q=4,6,8,10, \ldots$ and the solvent. The system is designed (see Sec. I and  \cite{li1998SM}) so that the molecules interact exclusively intermolecularly and the two interaction sites of the same $A$ molecule can not interact with two sites belonging to the same $B$ molecule. To formulate a
  mean-field theory for the system, we consider a lattice model (see Fig. 1 in the main text) where the $A$ molecules can occupy the bonds and $B$ molecules occupy the sites of the lattice. Solvent molecules, $S$, can occupy either the bonds or the sites as schematically shown in Fig. 1 in the paper. We define the total number of sites as $N_s$ and the total number of bonds as $N_b$ . For concreteness, we consider a square lattice and $q = 4$ for the B molecules (that is, a tetramer), however, the formalism is more general. For the square lattice, $N_b = 2N_s$ and the total number of bonds and sites in the lattice is $3N_s$. We consider that there are a total of $N_A^0$ $A$ molecules and $N_B^0$ $B$ molecules. Since all the sites and bonds of the lattice are occupied, conservation dictates that there must be $N_b-N_A^0=2N_s-N_A^0$ $S$ molecules on the bonds and $N_s-N_B^0$ $S$ molecules at the sites. As we discussed in the main text, the analysis of the phase separation involves two steps: (1) the $A$ and $B$ molecules associate with each other forming complexes, and (2) the complexes interact with the $B$ molecules leading to the phase separation. For tetramers, there can be four different complexes: $C_i$ with $i = 1, 2, 3, 4$ where $C_i$ represents a complex with $i$ distinct $A$ molecules being associated with a $B$ molecule as shown in Fig. 1. The
  concentration of free A molecules is $\rho_A = N_A/N_b$ where $N_A$ is the number of free $A$ molecules that are not associated with any $B$ molecules. Similarly, $\rho_B = N_B/N_s$ is the concentration of free $B$ molecules. This particular normalization accounts for the valence which is equal to $2 N_b/N_s$ since a bond is shared by two sites.  Since each of these dimensionless concentrations can be unity, we denote them by effective concentrations. The total concentrations of $A$ and $B$ molecules relative to the number of bonds and sites, respectively, are defined as $\rho^0_A$ and $\rho^0_B$. We define the concentrations of the $i$th complex as $\g_i$. 
  
  The $B$ molecules occupy the $N_s$ sites of the lattice and the $A$ molecules the $N_b$ bonds and  write  the entropic
  part of the total free energy per site as:
  \begin{align}\label{f_entropic}
  f_{entropic} =&2\rho_A\ln\rho_A+\rho_B\ln\rho_B+2(1-\rho_A^0)\ln(1-\rho_A^0) \nonumber\\
  &+(1-\rho_B^0)\ln(1-\rho_B^0)+\gamma_1\ln(4\g_1)+\g_2\ln(6\g_2) \nonumber\\
  &+\g_3\ln(4\g_3)+\g_4\ln\g_4 
  \end{align}
  where we have distinguished among the solvent molecules that are located at the bonds and at the sites, by the third and fourth terms of Eq. (\ref{f_entropic}). The numerical factors in the entropic contributions of the complexes come from the simple counting of the number of ways of associating the $A$ molecules with the $B$ molecules. We ignore the rotational contributions to the entropy. The attractive association of a $A$ molecule with a $B$ molecule reduces the free energy and $J_i$ is the reduction in free energy per total of sites and bonds associated with the formation of the $i$th complex. In our mean-field approximation, the complexes interact with the average number of $B$ molecules in the system, which can lead to phase separation.
  Then, the interaction part of the free energy per total of sites and bonds can be written as
  \begin{align}\label{f_interaction}
  f_{interaction}&=-J_1\g_1-J_2\g_2-J_3\g_3-J_4\g_4 \nonumber\\
  -(J-&J_{BB})\rho_B^0(\g_1+2\g_2+3\g_3+4\g_4)
  \end{align}
  In this expression we have used  a mean-field approximation for mathematical simplicity whereby the complexes (via the A molecules attached to them) interact with all the $B$ molecules on the sites. $J_{BB}$ is non-zero if the energy of associating a $B$ molecule with the interaction site of a $A$ molecule that already has a $B$ molecule associated with the other end of the $A$ dimer, compared with a $A$ molecule attached to a single $B$ is different. It appears that a good, semi-quantitative description of the experiments is obtained even if we set $J_{BB}=0$ (see main text and Discussion).
  
  Therefore, the total free energy per total of sites and bonds, $f=F/M$, where $F$ is the total free energy, can be written as sum of Eqs. (\ref{f_entropic}) and (\ref{f_interaction}) as
  \begin{align}\label{totf}
  f =&2\rho_A\ln\rho_A+\rho_B\ln\rho_B+2(1-\rho_A^0)\ln(1-\rho_A^0) \nonumber\\
  &+(1-\rho_B^0)\ln(1-\rho_B^0)+\gamma_1\ln(4\g_1)+\g_2\ln(6\g_2) \nonumber\\
  &+\g_3\ln(4\g_3)+\g_4\ln\g_4 \nonumber\\
  &-J_1\g_1-J_2\g_2-J_3\g_3-J_4\g_4 \nonumber\\
  &-(J-J_{BB})\rho_B^0(\g_1+2\g_2+3\g_3+4\g_4)
  \end{align}
  where $\rho_A=\rho_A^0-(\g_1+2\g_2+3\g_3+4\g_4)/2$ and $\rho_B=\rho_B^0-(\g_1+\g_2+\g_3+\g_4)$ are the concentrations of the interaction sites of free $A$ and free $B$ as we explained above.
  We first must minimize this free energy with respect to all of $\g_i$'s, which predicts the equilibrium concentrations of the complexes in terms of total concentrations of $A$ and $B$ molecules.

  A direct minimization to find the different $\g_i$'s is difficult, even numerically, because of the nonlinear equations arising as a result of the minimization. Therefore, we first cast them in an algebraically simpler form before solving them numerically. Minimizing the free energy, Eq. (\ref{totf}), with respect to $\g_i$, $i=1,2,3,4$, after a slight mathematical manipulation,leads to the following four equations
  \begin{subequations}
  	\begin{align}
  	- &\ln X-\ln Y+\ln\g_1=1+(J-\ln4)+(J-J_{BB})\rho_B^0 \label{eq1_min}\\
  	-&2\ln X-\ln Y+\ln\g_2=2+(2J-\ln6)+2(J-J_{BB})\rho_B^0\label{eq2_min}\\
  	-&3\ln X-\ln Y+\ln\g_3=3+(3J-\ln4)+3(J-J_{BB})\rho_B^0\label{eq3_min}\\
  	-&4\ln X-\ln Y+\ln\g_4=4+4J+4(J-J_{BB})\rho_B^0 \label{eq4_min}
  	\end{align}
  \end{subequations}
  where, we have defined $X=\rho_A=\rho_A^0-(\g_1+2\g_2+3\g_3+4\g_4)/2$ and $Y=\rho_B=\rho_B^0-(\g_1+\g_2+\g_3+\g_4)$ for the convenience of notation. We next subtract Eq. (\ref{eq1_min}) from Eq. (\ref{eq2_min}), Eq. (\ref{eq2_min}) from Eq. (\ref{eq3_min}) and Eq. (\ref{eq3_min}) from Eq. (\ref{eq4_min}) and define $\alpha=\exp[1+J+(J-J_{BB})\rho_B^0]$. These three equations, along with Eq. (\ref{eq1_min}) give
  \begin{subequations}\label{gamrel}
  	\begin{align}
  	\g_1=\f{XY\alpha}{4}\\
  	\f{\g_2}{X\g_1}=\f{2\alpha}{3}\\
  	\f{\g_3}{X\g_2}=\f{3\alpha}{2}\\
  	\f{\g_4}{X\g_3}=4\alpha.
  	\end{align}
  \end{subequations}
  From these equations, we obtain $\g_1=XY\alpha/4$, $\g_2=2\alpha X\g_1/3$, $\g_3=\alpha^2 X^2 \g_1$ and $\g_4=4\alpha^3X^3\g_1$. Replacing these relations back in the definitions of $X$ and $Y$ leads to
  \begin{subequations}\label{simplification1}
  	\begin{align}
  	X&=\rho_A^0-\left[1+\f{4\alpha X}{3}+3\alpha^2 X^2+16\alpha^3X^3\right]\f{XY\alpha}{8}\\
  	Y&=\rho_B^0-\left[1+\f{2\alpha X}{3}+\alpha^2X^2+4\alpha^3X^3\right]\f{\alpha X Y}{4}.
  	\end{align}
  \end{subequations}
  We now numerically solve these two equations for $X$ and $Y$ for given values of $\rho_A^0$ and $\rho_B^0$ and then find the $\g_i$'s from Eqs. (\ref{gamrel}). Since we will have to differentiate the free energy to find the binodal and spinodal, we use interpolation (we use the in-built function ListInterpolation of InterpolationOrder (3,3) of Mathematica \cite{mathematicaSM}) to obtain analytical (fitted) forms for the $\gamma_i$'s and insert those in the expression of the free energy, Eq. (\ref{totf}). Next we proceed through the usual procedure, as detailed below in Sec. \ref{sec_phasedia}, to obtain the phase diagrams for the system for a particular value of the interaction strength $J$ .

  \section{Calculation of the spinodal, binodal and critical points}
  \label{sec_phasedia}
  For the purpose of this section we set $J_{BB}=0$ and $J$ is the interaction strength and discuss below the effects of finite values of this interaction. Let us consider the free energy $f(\phi_A,\phi_B,J)$ of a three-component phase separating system where $\phi_A$ and $\phi_B$ are the concentrations of the two types of proteins in the single homogeneous phase and the solvent density is $(1-\phi_A-\phi_B)$. We want to calculate the spinodal line and the critical points at fixed $J$ and write $f(\phi_A,\phi_B,J)\equiv f(\phi_A,\phi_B)$. 
  
  To find the spinodal and critical points, we use the definitions that the spinodal is the limit of stability of the free energy up to quadratic order in small variations from the average concentrations and that at the critical points both the second and third order variations of the free energy in these small variations must be zero.  In general, the Taylor series expansion of the free energy about the average concentrations $\phi_A*0, \phi_B^0$ can be written up to third order in terms of $\delta\phi_A=\phi_A=-\phi_A^0$ and $\delta\phi_B=\phi_B=-\phi_B^0$:
  \begin{widetext}
  	\begin{align}
  	\delta f &= f(\phi_A,\phi_B) - f(\phi_A^0,\phi_B^0) \nonumber\\
  	&= f_A \delta\phi_A + f_B \delta\phi_B +\frac{1}{2} \left( f_{AA} \delta\phi_A^2 + 2 f_{AB}  \delta\phi_A \delta\phi_B + f_{BB} \delta\phi_B^2 \right)\nonumber \\  
  	&\hspace{1cm}+\frac{1}{6} \bigg( f_{AAA} \delta\phi_A^3 + 3 f_{AAB} \delta\phi_A^2 \delta\phi_B + 3 f_{ABB} \delta\phi_A \delta\phi_B^2 + f_{BBB}\delta\phi_B^3 \bigg)
  	\end{align}
  	To analyze this expansion in terms of the spinodal and critical points, it is useful to consider some fixed value of $\delta\phi_A$ and ask about the ratio of $n=\delta\phi_B/\delta\phi_A$.  This merely allows us to separate out a factor of $\delta\phi_A$ to an appropriate power in each term of the expansion which can now be written as:
  	
  	\begin{align}
  	\delta f = f(\phi_A,\phi_B) - f(\phi_A^0,\phi_B^0) =& \delta\phi_A  \left(  f_A + n \,f_B  \right)+ \frac{1}{2} \delta\phi_A^2 \left( f_{AA} + 2n \, f_{AB}   + n^2 \,  f_{BB}  \right) \nonumber  \\ &+\frac{1}{6}  \delta\phi_A^3 \left( f_{AAA}  + 3n \,  f_{AAB}  + 3n^2 \, f_{ABB}   +n^3 \,  f_{BBB}  \right)
  	\end{align}
  \end{widetext}
  The term linear in $\delta\phi_A$ is zero since the free energy (which includes the chemical potential) is a minimum.  The second term is a general quadratic function of $n$ and is positive definite as long as $f_{AA} f_{BB} – f_{AB}^2 >0$; the spinodal is the line in the plane defined by the two concentrations at which this inequality becomes an equality.  At the spinodal, we further require that the term quadratic in $\delta\phi_A$ in the expansion be equal to zero that determines that $n$ is given by $n= - f_{AB}/f_{BB} = - f_{AA}/ f_{AB}$.  Using this expression for $n$ in the third order term of the free energy, proportional to $\delta\phi_A^3$, we can find the condition at which this entire term is zero (i.e., the condition for the critical point): 
  \begin{align}\label{critpt}
  \f{\p^3f}{\p \phi_A^3}-&3 \f{f_{AB}}{f_{BB}} \f{\p^3f}{\p^2\phi_A\p \phi_B}+3\left(\f{f_{AB}}{f_{BB}}\right)^2\f{\p^3f}{\p \phi_A\p^2\phi_B} \nonumber\\
  &- \left(\f{f_{AB}}{f_{BB}}\right)^3 \f{\p^3f}{\p \phi_B^3}=0
  \end{align}
  where $f_{AB}\equiv \p^2f/\p \phi_A\p \phi_B$ etc.

  To obtain the binodal, we must look at the two phases; the dilute phase designated as $(\phi_A^{(1)},\phi_B^{(1)})$ and the dense phase, $(\phi_A^{(2)},\phi_B^{(2)})$. The chemical potentials of species $A$ and $B$ are given as
  \begin{equation}
  J_A=\f{\p f(\phi_A,\phi_B)}{\p \phi_A}; \,\,\, J_B=\f{\p f(\phi_A,\phi_B)}{\p \phi_B}
  \end{equation}
  and the osmotic pressure is given by
  \begin{equation}
  \Pi=f(\phi_A,\phi_B)-\phi_A \f{\p f(\phi_A,\phi_B)}{\p \phi_A}-\phi_B \f{\p f(\phi_A,\phi_B)}{\p \phi_B}
  \end{equation}
  In equilibrium, the chemical potentials of the two molecules in each phase as well as the osmotic pressures of the two phases must be equal. Thus, we have the following three conditions:
  \begin{subequations}\label{binodal_conditions}
  	\begin{align}
  	& \f{\p f(\phi_A,\phi_B)}{\p \phi_A}\bigg|_{\phi_A^{(1)},\phi_B^{(1)}}=\f{\p f(\phi_A,\phi_B)}{\p \phi_A}\bigg|_{\phi_A^{(2)},\phi_B^{(2)}} \\
  	& \f{\p f(\phi_A,\phi_B)}{\p \phi_B}\bigg|_{\phi_A^{(1)},\phi_B^{(1)}}=\f{\p f(\phi_A,\phi_B)}{\p \phi_B}\bigg|_{\phi_A^{(2)},\phi_B^{(2)}} \\
  	& \left(f(\phi_A,\phi_B)-\phi_A \f{\p f(\phi_A,\phi_B)}{\p \phi_A}-\phi_B \f{\p f(\phi_A,\phi_B)}{\p \phi_B}\right)_{\phi_A^{(1)},\phi_B^{(1)}} \nonumber\\
  	& \hspace{0.5cm} = \left(f(\phi_A,\phi_B)-\phi_A \f{\p f(\phi_A,\phi_B)}{\p \phi_A}-\phi_B \f{\p f(\phi_A,\phi_B)}{\p \phi_B}\right)_{\phi_A^{(2)},\phi_B^{(2)}}
  	\end{align}
  \end{subequations}
  For the binodal we have four variables, two densities in each of the dilute and dense phases, and we the  three equations \ref{binodal_conditions}.   We therefore take one variable as a parameter to be varied and  obtain the corresponding values of the other three variables using Eqs. (\ref{binodal_conditions}) and thus, obtain a line as shown in Fig. 2 in the paper.
  
  \section{Roles of the complexes}
  \label{role_complex}
  As we discussed in the main text, we must consider multi-particle interactions to model the experimental system. In order to include such interactions within a mean-field lattice model, we proceed in two separate stages. The $A$ and $B$ particles associate with each other to form the various complexes where different numbers of $A$ molecules can associate with the $B$ molecules. As shown in Fig. 1(b) in the main text, for solutions of tetramers, dimers and solvent, there are four possible complexes, $C_i$ with $i=1,2,3,4$, where  $i$ is the number of $A$ molecules being associated with a $B$ molecule.
  
  \begin{figure*}
  	\includegraphics[width=17cm]{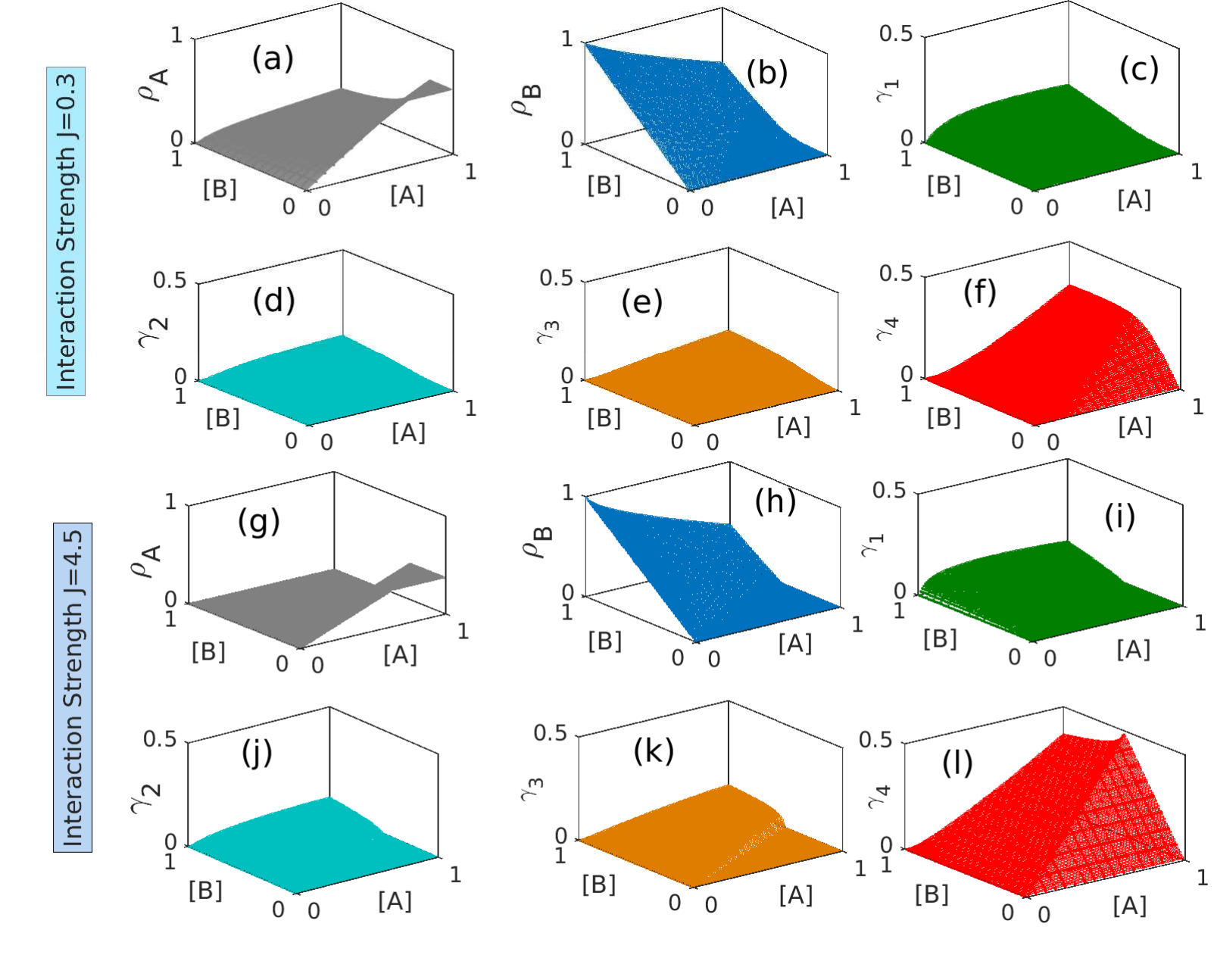}
  	\caption{Concentrations of free $A$ and $B$ proteins and the complexes for the system consisting of tetramers, dimers and solvent. Figs. (a)-(f) are for $J=0.3$ and (g)-(l) are for $J=4.5$. The system phase separates at around $J=3.0$. When $J$ is very small (e.g., 0.3 as in (a)-(f)), there is a significant fraction of free $A$ molecules even when there are enough $B$ molecules. However when the interaction is very strong ((g)-(l)) almost all of $A$ is efficiently used when there are enough $B$ molecules and the concentrations of the complexes are relatively large compared to when $J$ is small.}
  	\label{diffgam_diffmu}
  \end{figure*}

  We next examine the scenarios of small and large interaction strengths $J$ (relative to $k_BT$).
  In Fig. \ref{diffgam_diffmu} we show the concentrations of free $A$ and $B$ and those of the four complexes as functions of the concentrations of the $A$ and $B$ interaction sites (see above). Phase separation takes place for this system around $J\sim 3.0$; Figs. \ref{diffgam_diffmu}(a)-(f) are for $J=0.3$ and (g)-(l) for $J=4.5$. When the interaction is very small [Fig. \ref{diffgam_diffmu}(a)], there is a substantial amount of free $A$ in the solution even when there are sufficient $B$ molecules [Fig. \ref{diffgam_diffmu}(b)], to associate with them all, as expected. The concentrations of the complexes is also relatively small compared with the situation of very strong interactions as seen in Figs. \ref{diffgam_diffmu}(c-f) and \ref{diffgam_diffmu}(i-l) respectively. Fig. \ref{diffgam_diffmu}(g) and (h) shows that when the interaction is very strong, almost all of $A$ molecules are associated with $B$ molecules in complexes when there are sufficient $B$ molecules in the solution. The line showing the efficient use of $A$ molecules gives the symmetry line where phase separation is maximal - i.e., the tie line joining the two coexisting phases is longest. This line shows up in the formation of the complexes [Fig. \ref{diffgam_diffmu}] as well as in the binodal and spinodal phase diagrams. This is the point we have schematically alluded to through the two shaded regions, R1 and R2 in Fig. 1(a) in the paper.
  
  \section{Role of valency: comparison between two systems - hexamers, dimers and solvent vs. tetramers, dimers and solvent}
  To compare the role of valency in the phase separation, we consider two separate mixtures; one consisting of hexamers, dimers and solvent and the other of tetramers, dimers and solvent. For the mixture with hexamers, there can be six different complexes whose concentrations are denoted as $\g_i$ with $i=1, \ldots, 6$. We consider a cubic lattice, where the free energy $f_{hex}$ per overall number of lattice sites $M$ is
  \begin{align}
  f_{hex}&=3\rho_A\ln\rho_A+\rho_B\ln\rho_B+3(1-\rho_A^0)\ln(1-\rho_A^0) \nonumber\\
  &+(1-\rho_B^0)\ln(1-\rho_B^0)+\gamma_1\ln(6\g_1)+\g_2\ln(15\g_2) \nonumber\\
  &+\g_3\ln(20\g_3)+\g_4\ln(15\g_4)+\g_5\ln(6\g_5)+\g_6\ln\g_6 \nonumber\\
  &-(J-\ln6)\g_1-(2J-\ln15)\g_2-(3J-\ln20)\g_3 \nonumber\\
  &-(4J-\ln15)\g_4-(5J-\ln15)\g_5-6J\g_6 \nonumber\\
  &-J\rho_B^0(\g_1+2\g_2+3\g_3+4\g_4+5\g_5+6\g_6).
  \end{align}
  Using similar algebra as detailed in Sec. \ref{tetramerfreeenergy}, we obtain the concentrations for the complexes as
  $\g_1= XY\alpha/6$, $\g_2= 2\alpha X\g_1/5$, $\g_3= 3\alpha^2X^2\g_1/10$, $\g_4= 2\alpha^3X^3\g_1/5$, $\g_5= \alpha^4X^4\g_1$ and $\g_6= 6\alpha^5X^5\g_1$ along with 
  \begin{align}
  X &= \rho_A^0-(\g_1+2\g_2+3\g_3+4\g_4+5\g_5+6\g_6)/3 \nonumber\\
  Y &= \rho_B^0-(\g_1+\g_2+\g_3+\g_4+\g_5+\g_6).
  \end{align}
  
  The detailed expression of the free energy for the mixture of tetramers and the concentrations of the complexes are detailed in Sec. \ref{tetramerfreeenergy}.
  After a numerical solution of the equations for the concentrations of the complexes, we obtain their analytical forms via interpolation. Substituting these analytical expressions in the free energies, we numerically obtain the phase diagrams through the procedure detailed in Sec. \ref{numerics}. We have reported the phase diagrams in Fig. 3 in the main text of the paper.
  
  \section{Behavior of the minima of the phase diagram}
  To understand how the minima of the phase diagram vary with the interaction strength, we analyze a simple two-component system with a repulsive interaction of strength $J$ between the two different species (representing an attraction of each species to its own kind). The concentrations of the two components are given by $\phi$ and $(1-\phi)$ and the mean-field free energy of the system can be written \cite{sambookSM} as 
  \begin{equation}
  f=T[\phi\ln\phi+(1-\phi)\ln(1-\phi)]+\f{J}{2}\phi(1-\phi).
  \end{equation}
  This system has  symmetric phase diagram with the critical point at $\phi=1/2$. We consider the dilute side of the phase diagram (since the simple, two-component lattice gas is symmetric, the same behavior applies for the dense regime as well) and the chemical potential $J_{dilute}$ is given by
  \begin{equation}
  J_{dilute}\approx T(\ln\phi+1)+\f{J}{2}.
  \end{equation}
  From this expression, we see that $\phi\sim \exp(-J/2)$, where $J$ is the interaction strength, relative to $k_BT$. If we consider the three-component system, we also find (from our numerical results) that the minimum concentration varies exponentially with interaction strength. 
  
  \section{Details of numerical solution}
  \label{numerics}
  Solving the equations governing the phase diagram, as detailed in Sec. \ref{sec_phasedia}, is difficult even for a simple two-component system; one often needs to solve the equations numerically. Our theory is for a three-component system, but the complexity of the numerical solution increases by the formation of complexes. One first needs to minimize the free energy, $f$, with respect to the concentrations of complexes, $\gamma_i$, and solve the resulting equations simultaneously. Here we summarize the procedure for numerical solutions of the phase diagrams:
  \begin{enumerate}
  	\item 
  	We first analytically minimize $f$ and write $\gamma_i$ in such a way that we need  only solve two nonlinear equations simultaneously, as shown in Eq. (\ref{simplification1}).
  	\item 
  	In Mathematica \cite{mathematicaSM}, we numerically solve these equations and obtain the solutions for $\gamma_i$.
  	\item 
  	Via interpolation, we next obtain an analytical solution for $\gamma_i$, this is advantageous for the differentiation of $f$.
  	\item 
  	These analytical solutions for $\gamma_i$ are then used in the free energy and the equations for the spinodal, binodal and the critical points are solved numerically.
  \end{enumerate}
  
  As $J$ increases, the concentration of the dilute phase decreases exponentially. For the correct numerical solution, we therefore, have to take a finer grid. In our numerical solution, we have used a logarithmically spaced grid, but even then, the time required for the computations increases for larger values of  $J$, since one requires a finer grid. For example, with $q=4$ and $J=13$, the numerical solution for the binodal phase diagram close to the symmetry axis requires about 12 hours on a 3.7 GHz computer with 6-core i5 processors. The time requirement becomes larger for higher values of $q$ as more number of complexes increases the nonlinearity in Eq. (\ref{simplification1}).

  \section{Phase diagram along the symmetry axis}
\begin{figure}
  	\includegraphics[width=8.6cm]{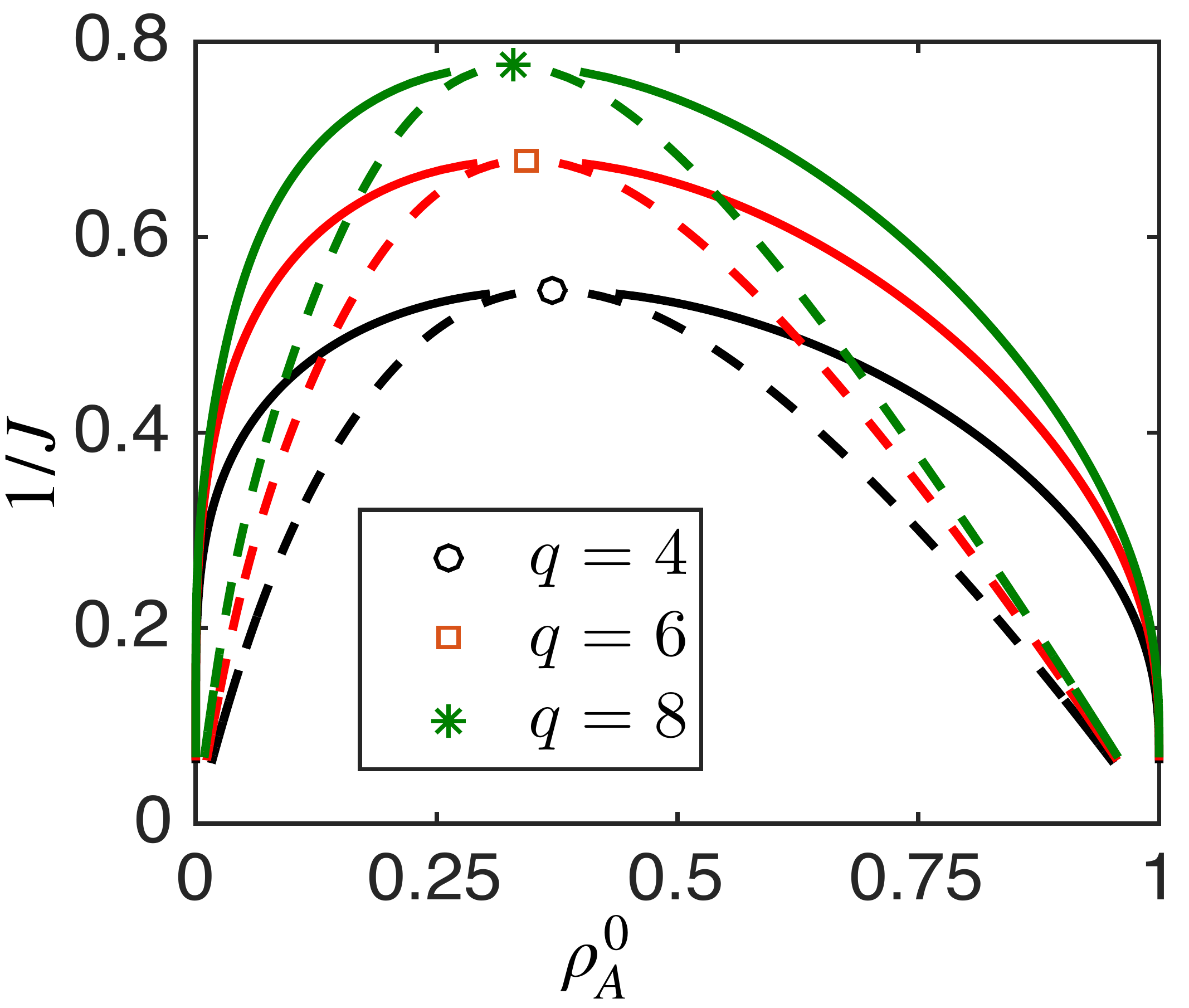}
  	\caption{Phase diagram along the symmetry axis for three systems where B species is either a tetramer, hexamer or octamer. Dotted lines are spinodals and continuous lines are binodals.}
  	\label{spinobino_sym}
  \end{figure}

  We have seen (Fig. 3a in the main text) that the maximum phase separation takes place along the symmetry axis, where $\rho_A^0=\rho_B^0$. Thus, solving the equations along this axis, by setting $\rho_B^0=\rho_A^0$, allows one to obtain $\Delta$ from this simplified model. Numerical solution of this model is much faster compared with computations based on the full model as the phase diagram becomes two-dimensional. The spinodal and binodal phase diagrams on this $(J,\rho_A^0)$ plane are presented in Fig. \ref{spinobino_sym}. We find that $\Delta$, obtained from this simplified model, agrees well with that obtained via the solution of the full three-dimensional model, as shown in Fig. 3(b).
  
  \section{Non-zero values of $J_{BB}$}
  We now  briefly comment on the role of the additional interaction, $J_{BB}$, in Eq. (\ref{f_interaction}). A value $J_{BB}>0$ means that associating a $B$ molecule with an $A$ molecule already engaged to another $B$ costs more energy compared with the association of a $B$ molecule with a free $A$. Such a penalty could reflect, for example, an allosteric communication between the two binding sites of an $A$ dimer. To minimize the number of parameters, we took $J_{BB}=0$ in the comparisons of theory and experiment. We point out that a small value of $J_{BB}$ does not introduce any qualitative differences in the phase separation scenario; however, large values of  $J_{BB}$ hinder phase separation even though complexes still form. Conversely, a negative value of $J_{BB}$, which could originate in avidity effects, would enhance phase separation. However, we assume that $J_{BB}$ which requires an interaction range equivalent to the length of the dimer, is smaller than $J$ which accounts for the more local binding of A and B.


\begin{thebibliography}{54}%
\makeatletter
\providecommand \@ifxundefined [1]{%
 \@ifx{#1\undefined}
}%
\providecommand \@ifnum [1]{%
 \ifnum #1\expandafter \@firstoftwo
 \else \expandafter \@secondoftwo
 \fi
}%
\providecommand \@ifx [1]{%
 \ifx #1\expandafter \@firstoftwo
 \else \expandafter \@secondoftwo
 \fi
}%
\providecommand \natexlab [1]{#1}%
\providecommand \enquote  [1]{``#1''}%
\providecommand \bibnamefont  [1]{#1}%
\providecommand \bibfnamefont [1]{#1}%
\providecommand \citenamefont [1]{#1}%
\providecommand \href@noop [0]{\@secondoftwo}%
\providecommand \href [0]{\begingroup \@sanitize@url \@href}%
\providecommand \@href[1]{\@@startlink{#1}\@@href}%
\providecommand \@@href[1]{\endgroup#1\@@endlink}%
\providecommand \@sanitize@url [0]{\catcode `\\12\catcode `\$12\catcode
  `\&12\catcode `\#12\catcode `\^12\catcode `\_12\catcode `\%12\relax}%
\providecommand \@@startlink[1]{}%
\providecommand \@@endlink[0]{}%
\providecommand \url  [0]{\begingroup\@sanitize@url \@url }%
\providecommand \@url [1]{\endgroup\@href {#1}{\urlprefix }}%
\providecommand \urlprefix  [0]{URL }%
\providecommand \Eprint [0]{\href }%
\providecommand \doibase [0]{http://dx.doi.org/}%
\providecommand \selectlanguage [0]{\@gobble}%
\providecommand \bibinfo  [0]{\@secondoftwo}%
\providecommand \bibfield  [0]{\@secondoftwo}%
\providecommand \translation [1]{[#1]}%
\providecommand \BibitemOpen [0]{}%
\providecommand \bibitemStop [0]{}%
\providecommand \bibitemNoStop [0]{.\EOS\space}%
\providecommand \EOS [0]{\spacefactor3000\relax}%
\providecommand \BibitemShut  [1]{\csname bibitem#1\endcsname}%
\let\auto@bib@innerbib\@empty
\bibitem [{\citenamefont {Zbinden}\ \emph {et~al.}(2020)\citenamefont
  {Zbinden}, \citenamefont {P{\'{e}}rez-Berlanga}, \citenamefont {Rossi},\ and\
  \citenamefont {Polymenidou}}]{zbinden2020}%
  \BibitemOpen
  \bibfield  {author} {\bibinfo {author} {\bibfnamefont {A.}~\bibnamefont
  {Zbinden}}, \bibinfo {author} {\bibfnamefont {M.}~\bibnamefont
  {P{\'{e}}rez-Berlanga}}, \bibinfo {author} {\bibfnamefont {P.~D.}\
  \bibnamefont {Rossi}}, \ and\ \bibinfo {author} {\bibfnamefont
  {M.}~\bibnamefont {Polymenidou}},\ }\href {\doibase
  10.1016/j.devcel.2020.09.014} {\bibfield  {journal} {\bibinfo  {journal}
  {Dev. Cell}\ }\textbf {\bibinfo {volume} {55}},\ \bibinfo {pages} {45}
  (\bibinfo {year} {2020})}\BibitemShut {NoStop}%
\bibitem [{\citenamefont {Peskett}\ \emph {et~al.}(2018)\citenamefont
  {Peskett}, \citenamefont {Rau}, \citenamefont {O’Driscoll}, \citenamefont
  {Patani}, \citenamefont {Lowe},\ and\ \citenamefont
  {R.Saibil}}]{peskett2018}%
  \BibitemOpen
  \bibfield  {author} {\bibinfo {author} {\bibfnamefont {T.~R.}\ \bibnamefont
  {Peskett}}, \bibinfo {author} {\bibfnamefont {F.}~\bibnamefont {Rau}},
  \bibinfo {author} {\bibfnamefont {J.}~\bibnamefont {O’Driscoll}}, \bibinfo
  {author} {\bibfnamefont {R.}~\bibnamefont {Patani}}, \bibinfo {author}
  {\bibfnamefont {A.~R.}\ \bibnamefont {Lowe}}, \ and\ \bibinfo {author}
  {\bibfnamefont {H.}~\bibnamefont {R.Saibil}},\ }\href {\doibase
  10.1016/j.molcel.2018.04.007} {\bibfield  {journal} {\bibinfo  {journal}
  {Mol. Cell}\ }\textbf {\bibinfo {volume} {70}},\ \bibinfo {pages} {588}
  (\bibinfo {year} {2018})}\BibitemShut {NoStop}%
\bibitem [{\citenamefont {Ghosh}\ \emph {et~al.}(2019)\citenamefont {Ghosh},
  \citenamefont {Sil}, \citenamefont {Dolai},\ and\ \citenamefont
  {Garai}}]{ghosh2019}%
  \BibitemOpen
  \bibfield  {author} {\bibinfo {author} {\bibfnamefont {S.}~\bibnamefont
  {Ghosh}}, \bibinfo {author} {\bibfnamefont {T.~B.}\ \bibnamefont {Sil}},
  \bibinfo {author} {\bibfnamefont {S.}~\bibnamefont {Dolai}}, \ and\ \bibinfo
  {author} {\bibfnamefont {K.}~\bibnamefont {Garai}},\ }\href {\doibase
  10.1111/febs.14988} {\bibfield  {journal} {\bibinfo  {journal} {FEBS J.}\
  }\textbf {\bibinfo {volume} {286}},\ \bibinfo {pages} {4737} (\bibinfo {year}
  {2019})}\BibitemShut {NoStop}%
\bibitem [{\citenamefont {Pytowski}\ \emph {et~al.}(2020)\citenamefont
  {Pytowski}, \citenamefont {Lee}, \citenamefont {Foley}, \citenamefont
  {Vaux},\ and\ \citenamefont {Jean}}]{pytowski2020}%
  \BibitemOpen
  \bibfield  {author} {\bibinfo {author} {\bibfnamefont {L.}~\bibnamefont
  {Pytowski}}, \bibinfo {author} {\bibfnamefont {C.~F.}\ \bibnamefont {Lee}},
  \bibinfo {author} {\bibfnamefont {A.~C.}\ \bibnamefont {Foley}}, \bibinfo
  {author} {\bibfnamefont {D.~J.}\ \bibnamefont {Vaux}}, \ and\ \bibinfo
  {author} {\bibfnamefont {L.}~\bibnamefont {Jean}},\ }\href {\doibase
  10.1073/pnas.1916716117} {\bibfield  {journal} {\bibinfo  {journal} {Proc.
  Natl. Acad. Sci. (USA)}\ }\textbf {\bibinfo {volume} {117}},\ \bibinfo
  {pages} {12050} (\bibinfo {year} {2020})}\BibitemShut {NoStop}%
\bibitem [{\citenamefont {Banani}\ \emph {et~al.}(2017)\citenamefont {Banani},
  \citenamefont {Lee}, \citenamefont {Hyman},\ and\ \citenamefont
  {Rosen}}]{banani2017}%
  \BibitemOpen
  \bibfield  {author} {\bibinfo {author} {\bibfnamefont {S.~F.}\ \bibnamefont
  {Banani}}, \bibinfo {author} {\bibfnamefont {H.~O.}\ \bibnamefont {Lee}},
  \bibinfo {author} {\bibfnamefont {A.~A.}\ \bibnamefont {Hyman}}, \ and\
  \bibinfo {author} {\bibfnamefont {M.~K.}\ \bibnamefont {Rosen}},\ }\href
  {\doibase 10.1038/nrm.2017.7} {\bibfield  {journal} {\bibinfo  {journal}
  {Nat. Rev. Mol. Cell Biol.}\ }\textbf {\bibinfo {volume} {18}},\ \bibinfo
  {pages} {285} (\bibinfo {year} {2017})}\BibitemShut {NoStop}%
\bibitem [{\citenamefont {Lyon}\ \emph {et~al.}(2021)\citenamefont {Lyon},
  \citenamefont {Peeples},\ and\ \citenamefont {Rosen}}]{lyon2021}%
  \BibitemOpen
  \bibfield  {author} {\bibinfo {author} {\bibfnamefont {A.~S.}\ \bibnamefont
  {Lyon}}, \bibinfo {author} {\bibfnamefont {W.~B.}\ \bibnamefont {Peeples}}, \
  and\ \bibinfo {author} {\bibfnamefont {M.~K.}\ \bibnamefont {Rosen}},\ }\href
  {\doibase 10.1038/ s41580-020-00303-z} {\bibfield  {journal} {\bibinfo
  {journal} {Nat. Rev.: Mol. Cell Biol.}\ }\textbf {\bibinfo {volume} {22}},\
  \bibinfo {pages} {215} (\bibinfo {year} {2021})}\BibitemShut {NoStop}%
\bibitem [{\citenamefont {Shin}\ and\ \citenamefont
  {Brangwynne}(2017)}]{shin2017}%
  \BibitemOpen
  \bibfield  {author} {\bibinfo {author} {\bibfnamefont {Y.}~\bibnamefont
  {Shin}}\ and\ \bibinfo {author} {\bibfnamefont {C.~P.}\ \bibnamefont
  {Brangwynne}},\ }\href {\doibase 10.1126/science.aaf4382} {\bibfield
  {journal} {\bibinfo  {journal} {Science}\ }\textbf {\bibinfo {volume}
  {357}},\ \bibinfo {pages} {eaaf4382} (\bibinfo {year} {2017})}\BibitemShut
  {NoStop}%
\bibitem [{\citenamefont {Berry}\ \emph {et~al.}(2018)\citenamefont {Berry},
  \citenamefont {Brangwynne},\ and\ \citenamefont {Haataja}}]{berry2018}%
  \BibitemOpen
  \bibfield  {author} {\bibinfo {author} {\bibfnamefont {J.}~\bibnamefont
  {Berry}}, \bibinfo {author} {\bibfnamefont {C.~P.}\ \bibnamefont
  {Brangwynne}}, \ and\ \bibinfo {author} {\bibfnamefont {M.}~\bibnamefont
  {Haataja}},\ }\href {\doibase 10.1088/1361-6633/aaa61e} {\bibfield  {journal}
  {\bibinfo  {journal} {Rep. Prog. Phys.}\ }\textbf {\bibinfo {volume} {81}},\
  \bibinfo {pages} {046601} (\bibinfo {year} {2018})}\BibitemShut {NoStop}%
\bibitem [{\citenamefont {Smith}\ \emph {et~al.}(2016)\citenamefont {Smith},
  \citenamefont {Calidas}, \citenamefont {Schmidt}, \citenamefont {Lu},
  \citenamefont {Rasoloson},\ and\ \citenamefont {Seydoux}}]{smith2016}%
  \BibitemOpen
  \bibfield  {author} {\bibinfo {author} {\bibfnamefont {J.}~\bibnamefont
  {Smith}}, \bibinfo {author} {\bibfnamefont {D.}~\bibnamefont {Calidas}},
  \bibinfo {author} {\bibfnamefont {H.}~\bibnamefont {Schmidt}}, \bibinfo
  {author} {\bibfnamefont {T.}~\bibnamefont {Lu}}, \bibinfo {author}
  {\bibfnamefont {D.}~\bibnamefont {Rasoloson}}, \ and\ \bibinfo {author}
  {\bibfnamefont {G.}~\bibnamefont {Seydoux}},\ }\href {\doibase
  10.7554/eLife.21337} {\bibfield  {journal} {\bibinfo  {journal} {eLife}\
  }\textbf {\bibinfo {volume} {5}},\ \bibinfo {pages} {e21337} (\bibinfo {year}
  {2016})}\BibitemShut {NoStop}%
\bibitem [{\citenamefont {Saha}\ \emph {et~al.}(2016)\citenamefont {Saha},
  \citenamefont {Weber}, \citenamefont {Nousch}, \citenamefont {Adame-Arana},
  \citenamefont {Hoege}, \citenamefont {Hein}, \citenamefont
  {Osborne-Nishimura}, \citenamefont {Mahamid}, \citenamefont {Jahnel},
  \citenamefont {Jawerth}, \citenamefont {Pozniakovski}, \citenamefont
  {Eckmann}, \citenamefont {J{\"{u}}licher},\ and\ \citenamefont
  {Hyman}}]{saha2016}%
  \BibitemOpen
  \bibfield  {author} {\bibinfo {author} {\bibfnamefont {S.}~\bibnamefont
  {Saha}}, \bibinfo {author} {\bibfnamefont {C.~A.}\ \bibnamefont {Weber}},
  \bibinfo {author} {\bibfnamefont {M.}~\bibnamefont {Nousch}}, \bibinfo
  {author} {\bibfnamefont {O.}~\bibnamefont {Adame-Arana}}, \bibinfo {author}
  {\bibfnamefont {C.}~\bibnamefont {Hoege}}, \bibinfo {author} {\bibfnamefont
  {M.~Y.}\ \bibnamefont {Hein}}, \bibinfo {author} {\bibfnamefont
  {E.}~\bibnamefont {Osborne-Nishimura}}, \bibinfo {author} {\bibfnamefont
  {J.}~\bibnamefont {Mahamid}}, \bibinfo {author} {\bibfnamefont
  {M.}~\bibnamefont {Jahnel}}, \bibinfo {author} {\bibfnamefont
  {L.}~\bibnamefont {Jawerth}}, \bibinfo {author} {\bibfnamefont
  {A.}~\bibnamefont {Pozniakovski}}, \bibinfo {author} {\bibfnamefont {C.~R.}\
  \bibnamefont {Eckmann}}, \bibinfo {author} {\bibfnamefont {F.}~\bibnamefont
  {J{\"{u}}licher}}, \ and\ \bibinfo {author} {\bibfnamefont {A.~A.}\
  \bibnamefont {Hyman}},\ }\href {\doibase 10.1016/j.cell.2016.08.006}
  {\bibfield  {journal} {\bibinfo  {journal} {Cell}\ }\textbf {\bibinfo
  {volume} {166}},\ \bibinfo {pages} {1} (\bibinfo {year} {2016})}\BibitemShut
  {NoStop}%
\bibitem [{\citenamefont {Protter}\ and\ \citenamefont
  {Parker}(2016)}]{protter2016}%
  \BibitemOpen
  \bibfield  {author} {\bibinfo {author} {\bibfnamefont {D.~S.~W.}\
  \bibnamefont {Protter}}\ and\ \bibinfo {author} {\bibfnamefont
  {R.}~\bibnamefont {Parker}},\ }\href {\doibase 10.1016/j.tcb.2016.05.004}
  {\bibfield  {journal} {\bibinfo  {journal} {Trends in Cell Biol.}\ }\textbf
  {\bibinfo {volume} {26}},\ \bibinfo {pages} {668} (\bibinfo {year}
  {2016})}\BibitemShut {NoStop}%
\bibitem [{\citenamefont {Kucherenko}\ and\ \citenamefont
  {Shcherbata}(2018)}]{kucherenko2018}%
  \BibitemOpen
  \bibfield  {author} {\bibinfo {author} {\bibfnamefont {M.~M.}\ \bibnamefont
  {Kucherenko}}\ and\ \bibinfo {author} {\bibfnamefont {H.~R.}\ \bibnamefont
  {Shcherbata}},\ }\href {\doibase 10.1038/s41467-017-02757-w} {\bibfield
  {journal} {\bibinfo  {journal} {Nat. Comm.}\ }\textbf {\bibinfo {volume}
  {9}},\ \bibinfo {pages} {312} (\bibinfo {year} {2018})}\BibitemShut {NoStop}%
\bibitem [{\citenamefont {Jayabalan}\ \emph {et~al.}(2016)\citenamefont
  {Jayabalan}, \citenamefont {Sanchez}, \citenamefont {Park}, \citenamefont
  {Yoon}, \citenamefont {Kang}, \citenamefont {Baek}, \citenamefont {Anderson},
  \citenamefont {Kee},\ and\ \citenamefont {Ohn}}]{jayabalan2016}%
  \BibitemOpen
  \bibfield  {author} {\bibinfo {author} {\bibfnamefont {A.~K.}\ \bibnamefont
  {Jayabalan}}, \bibinfo {author} {\bibfnamefont {A.}~\bibnamefont {Sanchez}},
  \bibinfo {author} {\bibfnamefont {R.~Y.}\ \bibnamefont {Park}}, \bibinfo
  {author} {\bibfnamefont {S.~P.}\ \bibnamefont {Yoon}}, \bibinfo {author}
  {\bibfnamefont {G.-Y.}\ \bibnamefont {Kang}}, \bibinfo {author}
  {\bibfnamefont {J.-H.}\ \bibnamefont {Baek}}, \bibinfo {author}
  {\bibfnamefont {P.}~\bibnamefont {Anderson}}, \bibinfo {author}
  {\bibfnamefont {Y.}~\bibnamefont {Kee}}, \ and\ \bibinfo {author}
  {\bibfnamefont {T.}~\bibnamefont {Ohn}},\ }\href {\doibase
  10.1038/ncomms12125} {\bibfield  {journal} {\bibinfo  {journal} {Nat. Comm.}\
  }\textbf {\bibinfo {volume} {7}},\ \bibinfo {pages} {12125} (\bibinfo {year}
  {2016})}\BibitemShut {NoStop}%
\bibitem [{\citenamefont {Seguin}\ \emph {et~al.}(2014)\citenamefont {Seguin},
  \citenamefont {Morelli}, \citenamefont {Vinet}, \citenamefont {Amore},
  \citenamefont {Biasi}, \citenamefont {Poletti}, \citenamefont {Rubinsztein},\
  and\ \citenamefont {Carra}}]{seguin2014}%
  \BibitemOpen
  \bibfield  {author} {\bibinfo {author} {\bibfnamefont {S.~J.}\ \bibnamefont
  {Seguin}}, \bibinfo {author} {\bibfnamefont {F.~F.}\ \bibnamefont {Morelli}},
  \bibinfo {author} {\bibfnamefont {J.}~\bibnamefont {Vinet}}, \bibinfo
  {author} {\bibfnamefont {D.}~\bibnamefont {Amore}}, \bibinfo {author}
  {\bibfnamefont {S.~D.}\ \bibnamefont {Biasi}}, \bibinfo {author}
  {\bibfnamefont {A.}~\bibnamefont {Poletti}}, \bibinfo {author} {\bibfnamefont
  {D.~C.}\ \bibnamefont {Rubinsztein}}, \ and\ \bibinfo {author} {\bibfnamefont
  {S.}~\bibnamefont {Carra}},\ }\href {\doibase 10.1038/cdd.2014.103}
  {\bibfield  {journal} {\bibinfo  {journal} {Cell Death and Diff.}\ }\textbf
  {\bibinfo {volume} {21}},\ \bibinfo {pages} {1838} (\bibinfo {year}
  {2014})}\BibitemShut {NoStop}%
\bibitem [{\citenamefont {Molliex}\ \emph {et~al.}(2015)\citenamefont
  {Molliex}, \citenamefont {Temirov}, \citenamefont {Lee}, \citenamefont
  {Coughlin}, \citenamefont {Kanagaraj}, \citenamefont {Kim}, \citenamefont
  {Mittag},\ and\ \citenamefont {Taylor}}]{molliex2015}%
  \BibitemOpen
  \bibfield  {author} {\bibinfo {author} {\bibfnamefont {A.}~\bibnamefont
  {Molliex}}, \bibinfo {author} {\bibfnamefont {J.}~\bibnamefont {Temirov}},
  \bibinfo {author} {\bibfnamefont {J.}~\bibnamefont {Lee}}, \bibinfo {author}
  {\bibfnamefont {M.}~\bibnamefont {Coughlin}}, \bibinfo {author}
  {\bibfnamefont {A.~P.}\ \bibnamefont {Kanagaraj}}, \bibinfo {author}
  {\bibfnamefont {H.~J.}\ \bibnamefont {Kim}}, \bibinfo {author} {\bibfnamefont
  {T.}~\bibnamefont {Mittag}}, \ and\ \bibinfo {author} {\bibfnamefont {J.~P.}\
  \bibnamefont {Taylor}},\ }\href {\doibase 10.1016/j.cell.2015.09.015}
  {\bibfield  {journal} {\bibinfo  {journal} {Cell}\ }\textbf {\bibinfo
  {volume} {163}},\ \bibinfo {pages} {123} (\bibinfo {year}
  {2015})}\BibitemShut {NoStop}%
\bibitem [{\citenamefont {Riback}\ \emph {et~al.}(2017)\citenamefont {Riback},
  \citenamefont {Katanski}, \citenamefont {Kear-Scott}, \citenamefont
  {Pilipenko}, \citenamefont {Rojek}, \citenamefont {Sosnick},\ and\
  \citenamefont {Drummond}}]{riback2017}%
  \BibitemOpen
  \bibfield  {author} {\bibinfo {author} {\bibfnamefont {J.~A.}\ \bibnamefont
  {Riback}}, \bibinfo {author} {\bibfnamefont {C.~D.}\ \bibnamefont
  {Katanski}}, \bibinfo {author} {\bibfnamefont {J.~L.}\ \bibnamefont
  {Kear-Scott}}, \bibinfo {author} {\bibfnamefont {E.~V.}\ \bibnamefont
  {Pilipenko}}, \bibinfo {author} {\bibfnamefont {A.~E.}\ \bibnamefont
  {Rojek}}, \bibinfo {author} {\bibfnamefont {T.~R.}\ \bibnamefont {Sosnick}},
  \ and\ \bibinfo {author} {\bibfnamefont {D.~A.}\ \bibnamefont {Drummond}},\
  }\href {\doibase 10.1016/j.cell.2017.02.027} {\bibfield  {journal} {\bibinfo
  {journal} {Cell}\ }\textbf {\bibinfo {volume} {168}},\ \bibinfo {pages}
  {1028} (\bibinfo {year} {2017})}\BibitemShut {NoStop}%
\bibitem [{\citenamefont {Strom}\ \emph {et~al.}(2017)\citenamefont {Strom},
  \citenamefont {Emelyanov}, \citenamefont {Mir}, \citenamefont {Fyodorov},
  \citenamefont {Darzacq},\ and\ \citenamefont {Karpen}}]{strom2017}%
  \BibitemOpen
  \bibfield  {author} {\bibinfo {author} {\bibfnamefont {A.~R.}\ \bibnamefont
  {Strom}}, \bibinfo {author} {\bibfnamefont {A.~V.}\ \bibnamefont
  {Emelyanov}}, \bibinfo {author} {\bibfnamefont {M.}~\bibnamefont {Mir}},
  \bibinfo {author} {\bibfnamefont {D.~V.}\ \bibnamefont {Fyodorov}}, \bibinfo
  {author} {\bibfnamefont {X.}~\bibnamefont {Darzacq}}, \ and\ \bibinfo
  {author} {\bibfnamefont {G.~H.}\ \bibnamefont {Karpen}},\ }\href {\doibase
  10.1038/nature22989} {\bibfield  {journal} {\bibinfo  {journal} {Nature}\
  }\textbf {\bibinfo {volume} {547}},\ \bibinfo {pages} {241} (\bibinfo {year}
  {2017})}\BibitemShut {NoStop}%
\bibitem [{\citenamefont {Mitrea}\ and\ \citenamefont
  {Kriwacki}(2016)}]{mitrea2016}%
  \BibitemOpen
  \bibfield  {author} {\bibinfo {author} {\bibfnamefont {D.~M.}\ \bibnamefont
  {Mitrea}}\ and\ \bibinfo {author} {\bibfnamefont {R.~W.}\ \bibnamefont
  {Kriwacki}},\ }\href {\doibase 10.1186/s12964-015-0125-7} {\bibfield
  {journal} {\bibinfo  {journal} {Cell Comm. and Signaling}\ }\textbf {\bibinfo
  {volume} {14}},\ \bibinfo {pages} {1} (\bibinfo {year} {2016})}\BibitemShut
  {NoStop}%
\bibitem [{\citenamefont {Liu}\ \emph {et~al.}(2020)\citenamefont {Liu},
  \citenamefont {Shen}, \citenamefont {Xie}, \citenamefont {Wei}, \citenamefont
  {Wong}, \citenamefont {Li}, \citenamefont {Zheng}, \citenamefont {Li},\ and\
  \citenamefont {Song}}]{liu2020}%
  \BibitemOpen
  \bibfield  {author} {\bibinfo {author} {\bibfnamefont {X.}~\bibnamefont
  {Liu}}, \bibinfo {author} {\bibfnamefont {J.}~\bibnamefont {Shen}}, \bibinfo
  {author} {\bibfnamefont {L.}~\bibnamefont {Xie}}, \bibinfo {author}
  {\bibfnamefont {Z.}~\bibnamefont {Wei}}, \bibinfo {author} {\bibfnamefont
  {C.}~\bibnamefont {Wong}}, \bibinfo {author} {\bibfnamefont {Y.}~\bibnamefont
  {Li}}, \bibinfo {author} {\bibfnamefont {X.}~\bibnamefont {Zheng}}, \bibinfo
  {author} {\bibfnamefont {P.}~\bibnamefont {Li}}, \ and\ \bibinfo {author}
  {\bibfnamefont {Y.}~\bibnamefont {Song}},\ }\href {\doibase
  10.1016/j.devcel.2019.11.019} {\bibfield  {journal} {\bibinfo  {journal}
  {Dev. Cell}\ }\textbf {\bibinfo {volume} {52}},\ \bibinfo {pages} {277}
  (\bibinfo {year} {2020})}\BibitemShut {NoStop}%
\bibitem [{\citenamefont {Quiroz}\ \emph {et~al.}(2020)\citenamefont {Quiroz},
  \citenamefont {Fiore}, \citenamefont {Levorse}, \citenamefont {Polak},
  \citenamefont {Wong}, \citenamefont {Pasolli},\ and\ \citenamefont
  {Fuchs}}]{quiroz2020}%
  \BibitemOpen
  \bibfield  {author} {\bibinfo {author} {\bibfnamefont {F.~G.}\ \bibnamefont
  {Quiroz}}, \bibinfo {author} {\bibfnamefont {V.~F.}\ \bibnamefont {Fiore}},
  \bibinfo {author} {\bibfnamefont {J.}~\bibnamefont {Levorse}}, \bibinfo
  {author} {\bibfnamefont {L.}~\bibnamefont {Polak}}, \bibinfo {author}
  {\bibfnamefont {E.}~\bibnamefont {Wong}}, \bibinfo {author} {\bibfnamefont
  {H.~A.}\ \bibnamefont {Pasolli}}, \ and\ \bibinfo {author} {\bibfnamefont
  {E.}~\bibnamefont {Fuchs}},\ }\href {\doibase 10.1126/science.aax9554}
  {\bibfield  {journal} {\bibinfo  {journal} {Science}\ }\textbf {\bibinfo
  {volume} {367}},\ \bibinfo {pages} {eaax9554} (\bibinfo {year}
  {2020})}\BibitemShut {NoStop}%
\bibitem [{\citenamefont {Zwicker}\ \emph {et~al.}(2014)\citenamefont
  {Zwicker}, \citenamefont {Decker}, \citenamefont {Jaensch}, \citenamefont
  {Hyman},\ and\ \citenamefont {J{\"{u}}licher}}]{zwicker2014}%
  \BibitemOpen
  \bibfield  {author} {\bibinfo {author} {\bibfnamefont {D.}~\bibnamefont
  {Zwicker}}, \bibinfo {author} {\bibfnamefont {M.}~\bibnamefont {Decker}},
  \bibinfo {author} {\bibfnamefont {S.}~\bibnamefont {Jaensch}}, \bibinfo
  {author} {\bibfnamefont {A.~A.}\ \bibnamefont {Hyman}}, \ and\ \bibinfo
  {author} {\bibfnamefont {F.}~\bibnamefont {J{\"{u}}licher}},\ }\href
  {\doibase 10.1073/pnas.1404855111} {\bibfield  {journal} {\bibinfo  {journal}
  {Proc. Natl. Acad. Sci. (USA)}\ }\textbf {\bibinfo {volume} {111}},\ \bibinfo
  {pages} {E2636} (\bibinfo {year} {2014})}\BibitemShut {NoStop}%
\bibitem [{\citenamefont {Zwicker}\ \emph {et~al.}(2015)\citenamefont
  {Zwicker}, \citenamefont {Hyman},\ and\ \citenamefont
  {J{\"{u}}licher}}]{zwicker2015}%
  \BibitemOpen
  \bibfield  {author} {\bibinfo {author} {\bibfnamefont {D.}~\bibnamefont
  {Zwicker}}, \bibinfo {author} {\bibfnamefont {A.~A.}\ \bibnamefont {Hyman}},
  \ and\ \bibinfo {author} {\bibfnamefont {F.}~\bibnamefont {J{\"{u}}licher}},\
  }\href {\doibase 10.1103/PhysRevE.92.012317} {\bibfield  {journal} {\bibinfo
  {journal} {Phys. Rev. E}\ }\textbf {\bibinfo {volume} {92}},\ \bibinfo
  {pages} {012317} (\bibinfo {year} {2015})}\BibitemShut {NoStop}%
\bibitem [{\citenamefont {Li}\ \emph {et~al.}(2012)\citenamefont {Li},
  \citenamefont {Banjade}, \citenamefont {Cheng}, \citenamefont {Kim},
  \citenamefont {Chen}, \citenamefont {Guo}, \citenamefont {Llaguno},
  \citenamefont {Hollingsworth}, \citenamefont {King}, \citenamefont {Banani},
  \citenamefont {Russo}, \citenamefont {Jiang}, \citenamefont {Nixon},\ and\
  \citenamefont {Rosen}}]{li2012}%
  \BibitemOpen
  \bibfield  {author} {\bibinfo {author} {\bibfnamefont {P.}~\bibnamefont
  {Li}}, \bibinfo {author} {\bibfnamefont {S.}~\bibnamefont {Banjade}},
  \bibinfo {author} {\bibfnamefont {H.-C.}\ \bibnamefont {Cheng}}, \bibinfo
  {author} {\bibfnamefont {S.}~\bibnamefont {Kim}}, \bibinfo {author}
  {\bibfnamefont {B.}~\bibnamefont {Chen}}, \bibinfo {author} {\bibfnamefont
  {L.}~\bibnamefont {Guo}}, \bibinfo {author} {\bibfnamefont {M.}~\bibnamefont
  {Llaguno}}, \bibinfo {author} {\bibfnamefont {J.~V.}\ \bibnamefont
  {Hollingsworth}}, \bibinfo {author} {\bibfnamefont {D.~S.}\ \bibnamefont
  {King}}, \bibinfo {author} {\bibfnamefont {S.~F.}\ \bibnamefont {Banani}},
  \bibinfo {author} {\bibfnamefont {P.~S.}\ \bibnamefont {Russo}}, \bibinfo
  {author} {\bibfnamefont {Q.-X.}\ \bibnamefont {Jiang}}, \bibinfo {author}
  {\bibfnamefont {B.~T.}\ \bibnamefont {Nixon}}, \ and\ \bibinfo {author}
  {\bibfnamefont {M.~K.}\ \bibnamefont {Rosen}},\ }\href {\doibase
  10.1038/nature10879} {\bibfield  {journal} {\bibinfo  {journal} {Nature}\
  }\textbf {\bibinfo {volume} {483}},\ \bibinfo {pages} {336} (\bibinfo {year}
  {2012})}\BibitemShut {NoStop}%
\bibitem [{\citenamefont {Klosin}\ \emph {et~al.}(2020)\citenamefont {Klosin},
  \citenamefont {Oltsch}, \citenamefont {Harmon}, \citenamefont {Honigmann},
  \citenamefont {J{\"{u}}licher}, \citenamefont {Hyman},\ and\ \citenamefont
  {Zechner}}]{klosin2020}%
  \BibitemOpen
  \bibfield  {author} {\bibinfo {author} {\bibfnamefont {A.}~\bibnamefont
  {Klosin}}, \bibinfo {author} {\bibfnamefont {F.}~\bibnamefont {Oltsch}},
  \bibinfo {author} {\bibfnamefont {T.}~\bibnamefont {Harmon}}, \bibinfo
  {author} {\bibfnamefont {A.}~\bibnamefont {Honigmann}}, \bibinfo {author}
  {\bibfnamefont {F.}~\bibnamefont {J{\"{u}}licher}}, \bibinfo {author}
  {\bibfnamefont {A.~A.}\ \bibnamefont {Hyman}}, \ and\ \bibinfo {author}
  {\bibfnamefont {C.}~\bibnamefont {Zechner}},\ }\href {\doibase
  10.1126/science.aav6691} {\bibfield  {journal} {\bibinfo  {journal}
  {Science}\ }\textbf {\bibinfo {volume} {367}},\ \bibinfo {pages} {464}
  (\bibinfo {year} {2020})}\BibitemShut {NoStop}%
\bibitem [{\citenamefont {Stoeger}\ \emph {et~al.}(2016)\citenamefont
  {Stoeger}, \citenamefont {Battich},\ and\ \citenamefont
  {Pelkmans}}]{stoger2016}%
  \BibitemOpen
  \bibfield  {author} {\bibinfo {author} {\bibfnamefont {T.}~\bibnamefont
  {Stoeger}}, \bibinfo {author} {\bibfnamefont {N.}~\bibnamefont {Battich}}, \
  and\ \bibinfo {author} {\bibfnamefont {L.}~\bibnamefont {Pelkmans}},\ }\href
  {\doibase 10.1016/j.cell.2016.02.005} {\bibfield  {journal} {\bibinfo
  {journal} {Cell}\ }\textbf {\bibinfo {volume} {164}},\ \bibinfo {pages}
  {1151} (\bibinfo {year} {2016})}\BibitemShut {NoStop}%
\bibitem [{\citenamefont {Deviri}\ and\ \citenamefont
  {Safran}(2021)}]{dan2021}%
  \BibitemOpen
  \bibfield  {author} {\bibinfo {author} {\bibfnamefont {D.}~\bibnamefont
  {Deviri}}\ and\ \bibinfo {author} {\bibfnamefont {S.~A.}\ \bibnamefont
  {Safran}},\ }\href {\doibase 10.1101/2021.01.05.425486} {\bibfield  {journal}
  {\bibinfo  {journal} {bioRxiv 2021.01.05.425486}\ } (\bibinfo {year}
  {2021}),\ 10.1101/2021.01.05.425486}\BibitemShut {NoStop}%
\bibitem [{\citenamefont {Prouteau}\ and\ \citenamefont
  {Loewith}(2018)}]{prouteau2018}%
  \BibitemOpen
  \bibfield  {author} {\bibinfo {author} {\bibfnamefont {M.}~\bibnamefont
  {Prouteau}}\ and\ \bibinfo {author} {\bibfnamefont {R.}~\bibnamefont
  {Loewith}},\ }\href {\doibase 10.3390/biom8040160} {\bibfield  {journal}
  {\bibinfo  {journal} {Biomolecules}\ }\textbf {\bibinfo {volume} {8}},\
  \bibinfo {pages} {160} (\bibinfo {year} {2018})}\BibitemShut {NoStop}%
\bibitem [{\citenamefont {Brangwynne}\ \emph {et~al.}(2009)\citenamefont
  {Brangwynne}, \citenamefont {Eckmann}, \citenamefont {Courson}, \citenamefont
  {Rybarska}, \citenamefont {Hoege}, \citenamefont {Gharakhani},\ and\
  \citenamefont {Hyman}}]{brangwynne2009}%
  \BibitemOpen
  \bibfield  {author} {\bibinfo {author} {\bibfnamefont {C.~P.}\ \bibnamefont
  {Brangwynne}}, \bibinfo {author} {\bibfnamefont {C.~R.}\ \bibnamefont
  {Eckmann}}, \bibinfo {author} {\bibfnamefont {D.~S.}\ \bibnamefont
  {Courson}}, \bibinfo {author} {\bibfnamefont {A.}~\bibnamefont {Rybarska}},
  \bibinfo {author} {\bibfnamefont {C.}~\bibnamefont {Hoege}}, \bibinfo
  {author} {\bibfnamefont {J.}~\bibnamefont {Gharakhani}}, \ and\ \bibinfo
  {author} {\bibfnamefont {F.~J. A.~A.}\ \bibnamefont {Hyman}},\ }\href
  {\doibase 10.1126/science.1172046} {\bibfield  {journal} {\bibinfo  {journal}
  {Science}\ }\textbf {\bibinfo {volume} {324}},\ \bibinfo {pages} {1729}
  (\bibinfo {year} {2009})}\BibitemShut {NoStop}%
\bibitem [{\citenamefont {Feric}\ \emph {et~al.}(2016)\citenamefont {Feric},
  \citenamefont {Vaidya}, \citenamefont {Harmon}, \citenamefont {Mitrea},
  \citenamefont {Zhu}, \citenamefont {Richardson}, \citenamefont {Kriwacki},
  \citenamefont {Pappu},\ and\ \citenamefont {Brangwynne}}]{feric2016}%
  \BibitemOpen
  \bibfield  {author} {\bibinfo {author} {\bibfnamefont {M.}~\bibnamefont
  {Feric}}, \bibinfo {author} {\bibfnamefont {N.}~\bibnamefont {Vaidya}},
  \bibinfo {author} {\bibfnamefont {T.~S.}\ \bibnamefont {Harmon}}, \bibinfo
  {author} {\bibfnamefont {D.~M.}\ \bibnamefont {Mitrea}}, \bibinfo {author}
  {\bibfnamefont {L.}~\bibnamefont {Zhu}}, \bibinfo {author} {\bibfnamefont
  {T.~M.}\ \bibnamefont {Richardson}}, \bibinfo {author} {\bibfnamefont
  {R.~W.}\ \bibnamefont {Kriwacki}}, \bibinfo {author} {\bibfnamefont {R.~V.}\
  \bibnamefont {Pappu}}, \ and\ \bibinfo {author} {\bibfnamefont {C.~P.}\
  \bibnamefont {Brangwynne}},\ }\href {\doibase 10.1016/j.cell.2016.04.047}
  {\bibfield  {journal} {\bibinfo  {journal} {Cell}\ }\textbf {\bibinfo
  {volume} {165}},\ \bibinfo {pages} {1686} (\bibinfo {year}
  {2016})}\BibitemShut {NoStop}%
\bibitem [{\citenamefont {Hyman}\ \emph {et~al.}(2014)\citenamefont {Hyman},
  \citenamefont {Weber},\ and\ \citenamefont {J{\"{u}}licher}}]{hyman2014}%
  \BibitemOpen
  \bibfield  {author} {\bibinfo {author} {\bibfnamefont {A.~A.}\ \bibnamefont
  {Hyman}}, \bibinfo {author} {\bibfnamefont {C.~A.}\ \bibnamefont {Weber}}, \
  and\ \bibinfo {author} {\bibfnamefont {F.}~\bibnamefont {J{\"{u}}licher}},\
  }\href {\doibase 10.1146/annurev-cellbio-100913-013325} {\bibfield  {journal}
  {\bibinfo  {journal} {Annu. Rev. Cell Dev. Biol.}\ }\textbf {\bibinfo
  {volume} {30}},\ \bibinfo {pages} {39} (\bibinfo {year} {2014})}\BibitemShut
  {NoStop}%
\bibitem [{\citenamefont {Alberti}\ \emph {et~al.}(2019)\citenamefont
  {Alberti}, \citenamefont {Gladfelter},\ and\ \citenamefont
  {Mittag}}]{alberti2019}%
  \BibitemOpen
  \bibfield  {author} {\bibinfo {author} {\bibfnamefont {S.}~\bibnamefont
  {Alberti}}, \bibinfo {author} {\bibfnamefont {A.}~\bibnamefont {Gladfelter}},
  \ and\ \bibinfo {author} {\bibfnamefont {T.}~\bibnamefont {Mittag}},\ }\href
  {\doibase 10.1016/j.cell.2018.12.035} {\bibfield  {journal} {\bibinfo
  {journal} {Cell}\ }\textbf {\bibinfo {volume} {176}},\ \bibinfo {pages} {419}
  (\bibinfo {year} {2019})}\BibitemShut {NoStop}%
\bibitem [{\citenamefont {Heidenreich}\ \emph {et~al.}(2020)\citenamefont
  {Heidenreich}, \citenamefont {Georgeson}, \citenamefont {Locatelli},
  \citenamefont {Rovigatti}, \citenamefont {Nandi}, \citenamefont {Steinberg},
  \citenamefont {Nadav}, \citenamefont {Shimoni}, \citenamefont {Safran},
  \citenamefont {Doye},\ and\ \citenamefont {Levy}}]{metapaper}%
  \BibitemOpen
  \bibfield  {author} {\bibinfo {author} {\bibfnamefont {M.}~\bibnamefont
  {Heidenreich}}, \bibinfo {author} {\bibfnamefont {J.}~\bibnamefont
  {Georgeson}}, \bibinfo {author} {\bibfnamefont {E.}~\bibnamefont
  {Locatelli}}, \bibinfo {author} {\bibfnamefont {L.}~\bibnamefont
  {Rovigatti}}, \bibinfo {author} {\bibfnamefont {S.~K.}\ \bibnamefont
  {Nandi}}, \bibinfo {author} {\bibfnamefont {A.}~\bibnamefont {Steinberg}},
  \bibinfo {author} {\bibfnamefont {Y.}~\bibnamefont {Nadav}}, \bibinfo
  {author} {\bibfnamefont {E.}~\bibnamefont {Shimoni}}, \bibinfo {author}
  {\bibfnamefont {S.}~\bibnamefont {Safran}}, \bibinfo {author} {\bibfnamefont
  {J.~P.~K.}\ \bibnamefont {Doye}}, \ and\ \bibinfo {author} {\bibfnamefont
  {E.~D.}\ \bibnamefont {Levy}},\ }\href {\doibase 10.1038/s41589-020-0576-z}
  {\bibfield  {journal} {\bibinfo  {journal} {Nat. Chem. Biol.}\ }\textbf
  {\bibinfo {volume} {16}},\ \bibinfo {pages} {939} (\bibinfo {year}
  {2020})}\BibitemShut {NoStop}%
\bibitem [{\citenamefont {Safran}(1994)}]{sambook}%
  \BibitemOpen
  \bibfield  {author} {\bibinfo {author} {\bibfnamefont {S.~A.}\ \bibnamefont
  {Safran}},\ }\href@noop {} {\emph {\bibinfo {title} {Statistical
  Thermodynamics of Surfaces, Interfaces, and Membranes}}}\ (\bibinfo
  {publisher} {Addison-Wesley Publishing Company},\ \bibinfo {year}
  {1994})\BibitemShut {NoStop}%
\bibitem [{\citenamefont {Dill}\ and\ \citenamefont
  {Bromberg}(2003)}]{dillbook}%
  \BibitemOpen
  \bibfield  {author} {\bibinfo {author} {\bibfnamefont {K.~A.}\ \bibnamefont
  {Dill}}\ and\ \bibinfo {author} {\bibfnamefont {S.}~\bibnamefont
  {Bromberg}},\ }\href@noop {} {\emph {\bibinfo {title} {Molecular Driving
  Forces: Statistical Thermodynamics in Chemistry and Biology}}}\ (\bibinfo
  {publisher} {Garland Science},\ \bibinfo {year} {2003})\BibitemShut {NoStop}%
\bibitem [{\citenamefont {Gennes}(1990)}]{degennesbook}%
  \BibitemOpen
  \bibfield  {author} {\bibinfo {author} {\bibfnamefont {P.~G.~D.}\
  \bibnamefont {Gennes}},\ }\href {\doibase 10.1017/CBO9780511569463} {\emph
  {\bibinfo {title} {Introduction to Polymer Dynamics}}}\ (\bibinfo
  {publisher} {Cambridge University Press},\ \bibinfo {year}
  {1990})\BibitemShut {NoStop}%
\bibitem [{\citenamefont {Flory}(1953)}]{florybook}%
  \BibitemOpen
  \bibfield  {author} {\bibinfo {author} {\bibfnamefont {P.~J.}\ \bibnamefont
  {Flory}},\ }\href@noop {} {\emph {\bibinfo {title} {Principles of Polymer
  Chemistry}}}\ (\bibinfo  {publisher} {Cornel University Press},\ \bibinfo
  {year} {1953})\BibitemShut {NoStop}%
\bibitem [{\citenamefont {Stockmayer}(1943)}]{stockmayer1943}%
  \BibitemOpen
  \bibfield  {author} {\bibinfo {author} {\bibfnamefont {W.~H.}\ \bibnamefont
  {Stockmayer}},\ }\href {\doibase 10.1063/1.1723803} {\bibfield  {journal}
  {\bibinfo  {journal} {J. Chem. Phys.}\ }\textbf {\bibinfo {volume} {11}},\
  \bibinfo {pages} {45} (\bibinfo {year} {1943})}\BibitemShut {NoStop}%
\bibitem [{\citenamefont {Stockmayer}(1944)}]{stockmayer1944}%
  \BibitemOpen
  \bibfield  {author} {\bibinfo {author} {\bibfnamefont {W.~H.}\ \bibnamefont
  {Stockmayer}},\ }\href {\doibase 10.1063/1.1723922} {\bibfield  {journal}
  {\bibinfo  {journal} {J. Chem. Phys.}\ }\textbf {\bibinfo {volume} {12}},\
  \bibinfo {pages} {125} (\bibinfo {year} {1944})}\BibitemShut {NoStop}%
\bibitem [{\citenamefont {Zilman}\ \emph {et~al.}(2003)\citenamefont {Zilman},
  \citenamefont {Kieffer}, \citenamefont {Molino}, \citenamefont {Porte},\ and\
  \citenamefont {Safran}}]{zilman2003}%
  \BibitemOpen
  \bibfield  {author} {\bibinfo {author} {\bibfnamefont {A.}~\bibnamefont
  {Zilman}}, \bibinfo {author} {\bibfnamefont {J.}~\bibnamefont {Kieffer}},
  \bibinfo {author} {\bibfnamefont {F.}~\bibnamefont {Molino}}, \bibinfo
  {author} {\bibfnamefont {G.}~\bibnamefont {Porte}}, \ and\ \bibinfo {author}
  {\bibfnamefont {S.~A.}\ \bibnamefont {Safran}},\ }\href {\doibase
  10.1103/PhysRevLett.91.015901} {\bibfield  {journal} {\bibinfo  {journal}
  {Phys. Rev. Lett.}\ }\textbf {\bibinfo {volume} {91}},\ \bibinfo {pages}
  {015901} (\bibinfo {year} {2003})}\BibitemShut {NoStop}%
\bibitem [{\citenamefont {Bianchi}\ \emph {et~al.}(2006)\citenamefont
  {Bianchi}, \citenamefont {Largo}, \citenamefont {Tartaglia}, \citenamefont
  {Zaccarelli},\ and\ \citenamefont {Sciortino}}]{bianchi2006}%
  \BibitemOpen
  \bibfield  {author} {\bibinfo {author} {\bibfnamefont {E.}~\bibnamefont
  {Bianchi}}, \bibinfo {author} {\bibfnamefont {J.}~\bibnamefont {Largo}},
  \bibinfo {author} {\bibfnamefont {P.}~\bibnamefont {Tartaglia}}, \bibinfo
  {author} {\bibfnamefont {E.}~\bibnamefont {Zaccarelli}}, \ and\ \bibinfo
  {author} {\bibfnamefont {F.}~\bibnamefont {Sciortino}},\ }\href {\doibase
  10.1103/PhysRevLett.97.168301} {\bibfield  {journal} {\bibinfo  {journal}
  {Phys. Rev. Lett.}\ }\textbf {\bibinfo {volume} {97}},\ \bibinfo {pages}
  {168301} (\bibinfo {year} {2006})}\BibitemShut {NoStop}%
\bibitem [{\citenamefont {Smallenburg}\ \emph {et~al.}(2013)\citenamefont
  {Smallenburg}, \citenamefont {Leibler},\ and\ \citenamefont
  {Sciortino}}]{smallenburg2013}%
  \BibitemOpen
  \bibfield  {author} {\bibinfo {author} {\bibfnamefont {F.}~\bibnamefont
  {Smallenburg}}, \bibinfo {author} {\bibfnamefont {L.}~\bibnamefont
  {Leibler}}, \ and\ \bibinfo {author} {\bibfnamefont {F.}~\bibnamefont
  {Sciortino}},\ }\href {\doibase 10.1103/PhysRevLett.111.188002} {\bibfield
  {journal} {\bibinfo  {journal} {Phys. Rev. Lett.}\ }\textbf {\bibinfo
  {volume} {111}},\ \bibinfo {pages} {188002} (\bibinfo {year}
  {2013})}\BibitemShut {NoStop}%
\bibitem [{\citenamefont {Wertheim}(1984{\natexlab{a}})}]{wertheimI}%
  \BibitemOpen
  \bibfield  {author} {\bibinfo {author} {\bibfnamefont {M.~S.}\ \bibnamefont
  {Wertheim}},\ }\href {\doibase 10.1007/BF01017362} {\bibfield  {journal}
  {\bibinfo  {journal} {J. Stat. Phys.}\ }\textbf {\bibinfo {volume} {35}},\
  \bibinfo {pages} {19} (\bibinfo {year} {1984}{\natexlab{a}})}\BibitemShut
  {NoStop}%
\bibitem [{\citenamefont {Wertheim}(1984{\natexlab{b}})}]{wertheimII}%
  \BibitemOpen
  \bibfield  {author} {\bibinfo {author} {\bibfnamefont {M.~S.}\ \bibnamefont
  {Wertheim}},\ }\href {\doibase 10.1007/BF01017363} {\bibfield  {journal}
  {\bibinfo  {journal} {J. Stat. Phys.}\ }\textbf {\bibinfo {volume} {35}},\
  \bibinfo {pages} {35} (\bibinfo {year} {1984}{\natexlab{b}})}\BibitemShut
  {NoStop}%
\bibitem [{\citenamefont {Wertheim}(1986{\natexlab{a}})}]{wertheimIII}%
  \BibitemOpen
  \bibfield  {author} {\bibinfo {author} {\bibfnamefont {M.~S.}\ \bibnamefont
  {Wertheim}},\ }\href {\doibase 10.1007/BF01127721} {\bibfield  {journal}
  {\bibinfo  {journal} {J. Stat. Phys.}\ }\textbf {\bibinfo {volume} {42}},\
  \bibinfo {pages} {459} (\bibinfo {year} {1986}{\natexlab{a}})}\BibitemShut
  {NoStop}%
\bibitem [{\citenamefont {Wertheim}(1986{\natexlab{b}})}]{wertheimIV}%
  \BibitemOpen
  \bibfield  {author} {\bibinfo {author} {\bibfnamefont {M.~S.}\ \bibnamefont
  {Wertheim}},\ }\href {\doibase 10.1007/BF01127722} {\bibfield  {journal}
  {\bibinfo  {journal} {J. Stat. Phys.}\ }\textbf {\bibinfo {volume} {42}},\
  \bibinfo {pages} {477} (\bibinfo {year} {1986}{\natexlab{b}})}\BibitemShut
  {NoStop}%
\bibitem [{\citenamefont {Harmon}\ \emph {et~al.}(2017)\citenamefont {Harmon},
  \citenamefont {Holehouse}, \citenamefont {Rosen},\ and\ \citenamefont
  {Pappu}}]{harmon2017}%
  \BibitemOpen
  \bibfield  {author} {\bibinfo {author} {\bibfnamefont {T.~S.}\ \bibnamefont
  {Harmon}}, \bibinfo {author} {\bibfnamefont {A.~S.}\ \bibnamefont
  {Holehouse}}, \bibinfo {author} {\bibfnamefont {M.~K.}\ \bibnamefont
  {Rosen}}, \ and\ \bibinfo {author} {\bibfnamefont {R.~V.}\ \bibnamefont
  {Pappu}},\ }\href {\doibase 10.7554/eLife.30294} {\bibfield  {journal}
  {\bibinfo  {journal} {eLife}\ }\textbf {\bibinfo {volume} {6}},\ \bibinfo
  {pages} {e30294} (\bibinfo {year} {2017})}\BibitemShut {NoStop}%
\bibitem [{\citenamefont {Harmon}\ \emph {et~al.}(2018)\citenamefont {Harmon},
  \citenamefont {Holehouse},\ and\ \citenamefont {Pappu}}]{harmon2018}%
  \BibitemOpen
  \bibfield  {author} {\bibinfo {author} {\bibfnamefont {T.~S.}\ \bibnamefont
  {Harmon}}, \bibinfo {author} {\bibfnamefont {A.~S.}\ \bibnamefont
  {Holehouse}}, \ and\ \bibinfo {author} {\bibfnamefont {R.~V.}\ \bibnamefont
  {Pappu}},\ }\href {\doibase 10.1088/1367-2630/aab8d9} {\bibfield  {journal}
  {\bibinfo  {journal} {New J. Phys.}\ }\textbf {\bibinfo {volume} {20}},\
  \bibinfo {pages} {045002} (\bibinfo {year} {2018})}\BibitemShut {NoStop}%
\bibitem [{\citenamefont {Choi}\ \emph {et~al.}(2019)\citenamefont {Choi},
  \citenamefont {Dar},\ and\ \citenamefont {Pappu}}]{choi2019}%
  \BibitemOpen
  \bibfield  {author} {\bibinfo {author} {\bibfnamefont {J.-M.}\ \bibnamefont
  {Choi}}, \bibinfo {author} {\bibfnamefont {F.}~\bibnamefont {Dar}}, \ and\
  \bibinfo {author} {\bibfnamefont {R.~V.}\ \bibnamefont {Pappu}},\ }\href
  {\doibase 10.1101/611095} {\bibfield  {journal} {\bibinfo  {journal}
  {BioarXiv}\ } (\bibinfo {year} {2019}),\ 10.1101/611095}\BibitemShut
  {NoStop}%
\bibitem [{\citenamefont {Wheeler}\ and\ \citenamefont
  {Andersen}(1980)}]{wheeler1980}%
  \BibitemOpen
  \bibfield  {author} {\bibinfo {author} {\bibfnamefont {J.~C.}\ \bibnamefont
  {Wheeler}}\ and\ \bibinfo {author} {\bibfnamefont {G.~R.}\ \bibnamefont
  {Andersen}},\ }\href {\doibase 10.1063/1.440061} {\bibfield  {journal}
  {\bibinfo  {journal} {J. Chem. Phys.}\ }\textbf {\bibinfo {volume} {73}},\
  \bibinfo {pages} {5778} (\bibinfo {year} {1980})}\BibitemShut {NoStop}%
\bibitem [{\citenamefont {Walker}\ and\ \citenamefont
  {Vause}(1980)}]{walker1980}%
  \BibitemOpen
  \bibfield  {author} {\bibinfo {author} {\bibfnamefont {J.~S.}\ \bibnamefont
  {Walker}}\ and\ \bibinfo {author} {\bibfnamefont {C.~A.}\ \bibnamefont
  {Vause}},\ }\href {\doibase 10.1016/0375-9601(80)90281-9} {\bibfield
  {journal} {\bibinfo  {journal} {Phys. Lett. A}\ }\textbf {\bibinfo {volume}
  {79}},\ \bibinfo {pages} {421} (\bibinfo {year} {1980})}\BibitemShut
  {NoStop}%
\bibitem [{\citenamefont {Li}\ \emph {et~al.}(1998)\citenamefont {Li},
  \citenamefont {Hamill}, \citenamefont {Hemmings}, \citenamefont {Moore},
  \citenamefont {James},\ and\ \citenamefont {Kleanthous}}]{li1998}%
  \BibitemOpen
  \bibfield  {author} {\bibinfo {author} {\bibfnamefont {W.}~\bibnamefont
  {Li}}, \bibinfo {author} {\bibfnamefont {S.~J.}\ \bibnamefont {Hamill}},
  \bibinfo {author} {\bibfnamefont {A.~M.}\ \bibnamefont {Hemmings}}, \bibinfo
  {author} {\bibfnamefont {G.~R.}\ \bibnamefont {Moore}}, \bibinfo {author}
  {\bibfnamefont {R.}~\bibnamefont {James}}, \ and\ \bibinfo {author}
  {\bibfnamefont {C.}~\bibnamefont {Kleanthous}},\ }\href {\doibase
  10.1021/bi9808621} {\bibfield  {journal} {\bibinfo  {journal} {Biochemistry}\
  }\textbf {\bibinfo {volume} {37}},\ \bibinfo {pages} {11771} (\bibinfo {year}
  {1998})}\BibitemShut {NoStop}%
\bibitem [{\citenamefont {Radhakrishnan}\ and\ \citenamefont
  {McConnel}(1999)}]{radhakrishnan1999}%
  \BibitemOpen
  \bibfield  {author} {\bibinfo {author} {\bibfnamefont {A.}~\bibnamefont
  {Radhakrishnan}}\ and\ \bibinfo {author} {\bibfnamefont {H.~M.}\ \bibnamefont
  {McConnel}},\ }\href {\doibase 10.1016/S0005-2736(03)00015-4} {\bibfield
  {journal} {\bibinfo  {journal} {Biophys. J.}\ }\textbf {\bibinfo {volume}
  {77}},\ \bibinfo {pages} {1507} (\bibinfo {year} {1999})}\BibitemShut
  {NoStop}%
\bibitem [{\citenamefont {Radhakrishnan}\ and\ \citenamefont
  {McConnel}(2005)}]{radhakrishnan2005}%
  \BibitemOpen
  \bibfield  {author} {\bibinfo {author} {\bibfnamefont {A.}~\bibnamefont
  {Radhakrishnan}}\ and\ \bibinfo {author} {\bibfnamefont {H.}~\bibnamefont
  {McConnel}},\ }\href {\doibase 10.1073/pnas.0506043102} {\bibfield  {journal}
  {\bibinfo  {journal} {Proc. Natl. Acad. of Sci. (USA)}\ }\textbf {\bibinfo
  {volume} {102}},\ \bibinfo {pages} {12662} (\bibinfo {year}
  {2005})}\BibitemShut {NoStop}%
\bibitem [{\citenamefont {Bray}(1994)}]{bray1994}%
  \BibitemOpen
  \bibfield  {author} {\bibinfo {author} {\bibfnamefont {A.~J.}\ \bibnamefont
  {Bray}},\ }\href {\doibase 10.1080/00018739400101505} {\bibfield  {journal}
  {\bibinfo  {journal} {Adv. Phys.}\ }\textbf {\bibinfo {volume} {43}},\
  \bibinfo {pages} {357} (\bibinfo {year} {1994})}\BibitemShut {NoStop}%
\end{thebibliography}

\begin{thebibliography}{4}%
 	\makeatletter
 	\providecommand \@ifxundefined [1]{%
 		\@ifx{#1\undefined}
 	}%
 	\providecommand \@ifnum [1]{%
 		\ifnum #1\expandafter \@firstoftwo
 		\else \expandafter \@secondoftwo
 		\fi
 	}%
 	\providecommand \@ifx [1]{%
 		\ifx #1\expandafter \@firstoftwo
 		\else \expandafter \@secondoftwo
 		\fi
 	}%
 	\providecommand \natexlab [1]{#1}%
 	\providecommand \enquote  [1]{``#1''}%
 	\providecommand \bibnamefont  [1]{#1}%
 	\providecommand \bibfnamefont [1]{#1}%
 	\providecommand \citenamefont [1]{#1}%
 	\providecommand \href@noop [0]{\@secondoftwo}%
 	\providecommand \href [0]{\begingroup \@sanitize@url \@href}%
 	\providecommand \@href[1]{\@@startlink{#1}\@@href}%
 	\providecommand \@@href[1]{\endgroup#1\@@endlink}%
 	\providecommand \@sanitize@url [0]{\catcode `\\12\catcode `\$12\catcode
 		`\&12\catcode `\#12\catcode `\^12\catcode `\_12\catcode `\%12\relax}%
 	\providecommand \@@startlink[1]{}%
 	\providecommand \@@endlink[0]{}%
 	\providecommand \url  [0]{\begingroup\@sanitize@url \@url }%
 	\providecommand \@url [1]{\endgroup\@href {#1}{\urlprefix }}%
 	\providecommand \urlprefix  [0]{URL }%
 	\providecommand \Eprint [0]{\href }%
 	\providecommand \doibase [0]{http://dx.doi.org/}%
 	\providecommand \selectlanguage [0]{\@gobble}%
 	\providecommand \bibinfo  [0]{\@secondoftwo}%
 	\providecommand \bibfield  [0]{\@secondoftwo}%
 	\providecommand \translation [1]{[#1]}%
 	\providecommand \BibitemOpen [0]{}%
 	\providecommand \bibitemStop [0]{}%
 	\providecommand \bibitemNoStop [0]{.\EOS\space}%
 	\providecommand \EOS [0]{\spacefactor3000\relax}%
 	\providecommand \BibitemShut  [1]{\csname bibitem#1\endcsname}%
 	\let\auto@bib@innerbib\@empty
 	\bibitem [{\citenamefont {Heidenreich}\ \emph {et~al.}(2020)\citenamefont
 		{Heidenreich}, \citenamefont {Georgeson}, \citenamefont {Locatelli},
 		\citenamefont {Rovigatti}, \citenamefont {Nandi}, \citenamefont {Steinberg},
 		\citenamefont {Nadav}, \citenamefont {Shimoni}, \citenamefont {Safran},
 		\citenamefont {Doye},\ and\ \citenamefont {Levy}}]{metapaperSM}%
 	\BibitemOpen
 	\bibfield  {author} {\bibinfo {author} {\bibfnamefont {M.}~\bibnamefont
 			{Heidenreich}}, \bibinfo {author} {\bibfnamefont {J.}~\bibnamefont
 			{Georgeson}}, \bibinfo {author} {\bibfnamefont {E.}~\bibnamefont
 			{Locatelli}}, \bibinfo {author} {\bibfnamefont {L.}~\bibnamefont
 			{Rovigatti}}, \bibinfo {author} {\bibfnamefont {S.~K.}\ \bibnamefont
 			{Nandi}}, \bibinfo {author} {\bibfnamefont {A.}~\bibnamefont {Steinberg}},
 		\bibinfo {author} {\bibfnamefont {Y.}~\bibnamefont {Nadav}}, \bibinfo
 		{author} {\bibfnamefont {E.}~\bibnamefont {Shimoni}}, \bibinfo {author}
 		{\bibfnamefont {S.}~\bibnamefont {Safran}}, \bibinfo {author} {\bibfnamefont
 			{J.~P.~K.}\ \bibnamefont {Doye}}, \ and\ \bibinfo {author} {\bibfnamefont
 			{E.~D.}\ \bibnamefont {Levy}},\ }\href {\doibase 10.1038/s41589-020-0576-z}
 	{\bibfield  {journal} {\bibinfo  {journal} {Nat. Chem. Biol.}\ }\textbf
 		{\bibinfo {volume} {16}},\ \bibinfo {pages} {939} (\bibinfo {year}
 		{2020})}\BibitemShut {NoStop}%
 	\bibitem [{\citenamefont {Li}\ \emph {et~al.}(1998)\citenamefont {Li},
 		\citenamefont {Hamill}, \citenamefont {Hemmings}, \citenamefont {Moore},
 		\citenamefont {James},\ and\ \citenamefont {Kleanthous}}]{li1998SM}%
 	\BibitemOpen
 	\bibfield  {author} {\bibinfo {author} {\bibfnamefont {W.}~\bibnamefont
 			{Li}}, \bibinfo {author} {\bibfnamefont {S.~J.}\ \bibnamefont {Hamill}},
 		\bibinfo {author} {\bibfnamefont {A.~M.}\ \bibnamefont {Hemmings}}, \bibinfo
 		{author} {\bibfnamefont {G.~R.}\ \bibnamefont {Moore}}, \bibinfo {author}
 		{\bibfnamefont {R.}~\bibnamefont {James}}, \ and\ \bibinfo {author}
 		{\bibfnamefont {C.}~\bibnamefont {Kleanthous}},\ }\href {\doibase
 		10.1021/bi9808621} {\bibfield  {journal} {\bibinfo  {journal} {Biochemistry}\
 		}\textbf {\bibinfo {volume} {37}},\ \bibinfo {pages} {11771} (\bibinfo {year}
 		{1998})}\BibitemShut {NoStop}%
 	\bibitem [{\citenamefont {{Wolfram Research{,} Inc}}(2016)}]{mathematicaSM}%
 	\BibitemOpen
 	\bibfield  {author} {\bibinfo {author} {\bibnamefont {{Wolfram Research{,}
 					Inc}}},\ }\href@noop {} {\enquote {\bibinfo {title} {Mathematica, {V}ersion
 				11.0},}\ } (\bibinfo {year} {2016}),\ \bibinfo {note} {wolfram Research,
 		Inc., Champaign, Illinois}\BibitemShut {NoStop}%
 	\bibitem [{\citenamefont {Safran}(1994)}]{sambookSM}%
 	\BibitemOpen
 	\bibfield  {author} {\bibinfo {author} {\bibfnamefont {S.~A.}\ \bibnamefont
 			{Safran}},\ }\href@noop {} {\emph {\bibinfo {title} {Statistical
 				Thermodynamics of Surfaces, Interfaces, and Membranes}}}\ (\bibinfo
 	{publisher} {Addison-Wesley Publishing Company},\ \bibinfo {year}
 	{1994})\BibitemShut {NoStop}%
 \end{thebibliography}
\end{document}